\documentclass[aip,jcp,reprint]{revtex4-1}
\usepackage[dvips]{graphicx}
\usepackage{amsmath}
\usepackage{bm}
\usepackage{amsfonts}
\usepackage{wasysym}
\usepackage[table]{xcolor}
\usepackage{array}
\usepackage{hyperref}
\usepackage{tikz}
\usepackage{soul}

\newcommand{\esc}{\!\cdot\!}
\usepackage{color}

\begin{document}

\title{Finite element discretization of non-linear diffusion  equations with thermal fluctuations}

\author{J.A. de la Torre}
\author{Pep Espa{\~n}ol}
\affiliation{Departamento  de F\'{\i}sica Fundamental,  UNED, Apartado
60141, 28080 Madrid, Spain}

\author{Aleksandar Donev}

\affiliation{Courant  Institute  of  Mathematical Sciences,  New  York
  University, 251 Mercer Street, 10012 NY, USA}

\date{21st October 2014}


\begin{abstract}
  We present a finite element discretization of a non-linear diffusion
  equation used in the field of critical phenomena and, more recently,
  in  the   context  of   Dynamic  Density  Functional   Theory.   The
  discretized  equation  preserves  the  structure  of  the  continuum
  equation. Specifically,  it conserves the total  number of particles
  and  fulfills  an H-theorem  as  the  original partial  differential
  equation.   The   discretization  proposed  suggests   a  particular
  definition  of the  discrete hydrodynamic  variables in  microscopic
  terms.  These variables are then used  to obtain, with the theory of
  coarse-graining,  their  dynamic  equations for  both  averages  and
  fluctuations.  The  hydrodynamic variables defined in  this way lead
  to  microscopically  derived  hydrodynamic  equations  that  have  a
  natural  interpretation  in  terms of  discretization  of  continuum
  equations.  Also,  the theory  of coarse-graining allows  to discuss
  the introduction  of thermal  fluctuations in a  physically sensible
  way.   The  methodology proposed  for  the  introduction of  thermal
  fluctuations in finite element methods is general and valid for both
  regular and irregular grids in  arbitrary dimensions.  We focus here
  on simulations  of the Ginzburg-Landau free  energy functional using
  both regular and  irregular 1D grids.  Convergence  of the numerical
  results is obtained for the  static and dynamic structure factors as
  the resolution of the grid is increased.
 \end{abstract}

 \maketitle

\section{Introduction}

The description  of transport processes  in soft matter  usually makes
use of  partial differential equations that  are typically non-linear.
The equilibrium  properties of  the system are  described with  a free
energy functional  and the transport properties  are described through
conservation equations. A typical example of such partial differential
equations (PDEs) is
\begin{align}
  \partial_t c({\bf r},t) &=\boldsymbol{\nabla}\esc \left[\Gamma(c({\bf r},t))
\boldsymbol{\nabla}\frac{\delta {\cal F}}{\delta c({\bf r})}[c]\right]
\label{PDE}
\end{align}
that governs the  dynamics of the concentration  field $c({\bf r},t)$.
Eq. (\ref{PDE}) has  become  the focus  of Dynamic  Density Functional
Theory  (DDFT) for  the  study of  dynamics  of colloidal  suspensions
\cite{Kawasaki1994,Marconi1999,Archer2004,Leonard2013}.

The two  quantities that  enter this
equation are the free energy functional ${\cal F}[c]$ and the mobility
coefficient  $\Gamma(c)$   that  may   depend,  in  general,   on  the
concentration   field.   The   partial  differential   equation  (PDE)
(\ref{PDE}) is paradigmatic in that it captures two essential features
of a non  equilibrium system.  On one hand, being  in divergence form,
Eq.  (\ref{PDE})  conserves  the  number of  particles  $N=\int  d{\bf
  r}c({\bf r},t)$.  On the other, it fulfill an H-theorem because the
time derivative of ${\cal F}[c]$  is always negative provided that the
mobility $\Gamma(c)$ is  positive.

Fluctuations are also relevant for soft matter and they are important
when  Brownian motion,  critical phenomena,  transitions events,  etc.
are  of  interest. Since  the  seminal  work  by Landau  and  Lifshitz
\cite{Landau1959},  thermal fluctuations  in  a  conservative PDE  are
introduced phenomenologically  through the divergence of  a stochastic
flux.  For  the particular example  of the above  non-linear diffusion
equation, the stochastic partial  differential equation (SPDE) has the
form
\begin{align}
  \partial_t c({\bf r},t) &=\boldsymbol{\nabla}\esc
\left[\Gamma(c({\bf r},t))
\boldsymbol{\nabla}\frac{\delta {\cal F}}{\delta c({\bf r})}[c]\right]
+\boldsymbol{\nabla}\esc\tilde{\bf J}({\bf r},t)
\label{SPDE}
\end{align}
where the stochastic mass flux $ \tilde{\bf J}({\bf r},t)$ is given by
\begin{align}
 \tilde{\bf J}({\bf r},t)&=
\sqrt{2 k_B T \Gamma}\;\boldsymbol{\zeta}({\bf r},t)
\label{J}
\end{align}
which obviously requires that $\Gamma>0$, and $\boldsymbol{\zeta}({\bf
  r},t)$  is a  white noise  in space  and time.  We will  discuss the
stochastic  interpretation   of  Eq.  (\ref{SPDE})  later   on.   This
stochastic  term ensures  that the  functional Fokker-Planck  Equation
equivalent  to  (\ref{SPDE}) has  formally  as  invariant measure  the
canonical equilibrium functional probability distribution
\begin{align}
  P^{\rm eq}[c] &=\frac{1}{Z}\exp\{-{\cal F}[c]/k_BT\}
\nonumber\\
Z&=\int {\cal D}c \exp\{-{\cal F}[c]/k_BT\}
\label{Peq}
\end{align}
where   the  partition   function  $Z$   normalizes  the   probability
distribution.   Fluctuating equations  of the form  (\ref{SPDE}) have
been  considered in  the DDFT  literature \cite{Leonard2013},  where a
debate   on    its   physical    meaning   has   arisen    (see   Ref.
\cite{Archer2004a} for a review).  Eq.  (\ref{SPDE}) has been used for
the  description  of  phase separation  \cite{Bray1994}  and  critical
phenomena, where  it is known  as Model B  in the terminology  of Ref.
\cite{Hohenberg1977}.

Despite  the formal  similarity between  (\ref{PDE}) and  (\ref{SPDE})
they are  very different kinds of  equations, not only because  one is
deterministic  and  the  other  stochastic.  As  we  will  discuss  in
Sec. \ref{Sec:ToCG}, the symbols  in Eq.  (\ref{PDE}) and (\ref{SPDE})
need to have  different physical meaning.  From  a purely mathematical
point of view, the very existence  of an equation like (\ref{SPDE}) or
a functional  like (\ref{Peq}) is  a delicate  point, due to  the fact
that the  noise $\boldsymbol{\zeta}({\bf  r},t)$ and the  field itself
$c({\bf r},t)$ are very irregular objects \cite{Ryser2012,Hairer2012}.
For example, in  the Ginzburg-Landau free energy  model, the partition
function $Z$ in Eq. (\ref{Peq}) has a proper continuum limit in 1D but
it is divergent in $D>1$ due to the so called ultraviolet catastrophe.
In this latter  case, renormalization group techniques  have been used
in       order      to       recover      a       continuum      limit
\cite{Benzi1989,Bettencourt1999,Gagne2000,Lythe2001,Cassol-Seewald2012}.
A rigorous mathematical analysis of  the renormalization of SPDEs near
the     critical     point     has     been     conducted     recently
\cite{Hairer2014,HairerReview}.  Alternatively, one may regularize the
equation by  introducing a physical coarse-graining  length.  This may
take the  form of regularization  of the noise, e.g.   replacing white
noise  with  colored  noise,  or  regularization  of  nonlinear  terms
\cite{DiffusionJSTAT}.

From a  computational point of view,  the numerical solution
  of partial  differential equations like (\ref{PDE})  always requires
  to  convert the  problem  in continuum  space into  a  problem in  a
  discrete  space,  amenable  of  treatment with  a  computer.   Usual
  procedures for discretization rely on assigning values of the fields
  to nodes  of a grid.  We  are interested in discretizations  on {\em
    arbitrary} (not necessarily regular) grids because arbitrary grids
  can accommodate  complex geometries  and allow for  adaptive spatial
  resolution.  Traditionally,  the numerical solution of  SPDEs of the
  kind  (\ref{SPDE})  have  resorted   to  finite  difference  schemes
  \cite{Lythe2001,Garcia-Ojalvo1999},  that are  easy to  implement in
  regular lattices.   Strictly speaking,  though, a  finite difference
  scheme  for an  SPDE like  (\ref{SPDE}) (without  regularization) is
  meaningless in higher dimensions because taking the point-wise value
  of the  field is  not appropriate.   Instead, one  can use  a finite
  volume  method,  in which  the  discrete  variables are  the  fields
  integrated  over the  cell volume  \cite{Donev2010}.  The  resulting
  algorithm in regular grids looks like a finite difference method but
  the variables  have very  different meanings.  While  finite volumes
  may   deal   with   adaptive   resolutions   and   irregular   grids
  \cite{Donev2010},  finite  elements  are  often  most  natural  when
  considering complicated boundary conditions.  Finite element methods
  for  the  solution  of  SPDEs  are just  beginning  to  be  explored
  \cite{Walsh2005,Yan2005,Krein2009,Plunkett}.

In  this  work,   we  present  a  finite   element  discretization  of
(\ref{PDE})  that  captures the  two  essential  ingredients of  exact
conservation and  fulfillment of the  Second Law,  and can be  used in
arbitrary grids.  While this may be regarded as a standard exercise in
numerical analysis, it  is a preliminary step for the  formulation of a
finite element discretization  of an SPDE like  (\ref{SPDE}). We point
out that  an equation like  (\ref{SPDE}) requires an  understanding of
its  microscopic underpinning  and that  (\ref{SPDE}) \textit{is  not}
just ``Eq. (\ref{PDE}) with  added thermal fluctuations''.  

Equations  (\ref{PDE})  and  (\ref{SPDE})   have  been  used  for  the
description of  colloidal suspensions out  of equilibrium, and  in the
study  of  critical  phenomena  of fluids.   In  these  fields,  these
equations correspond  to coarse-grained  (CG) descriptions  of systems
that  at  a  microscopic  level  are made  of  particles  governed  by
Hamilton's equations.  This  is distinct from what  happens in Quantum
Field  Theory  for  which  similar   equations  are  regarded  as  the
fundamental starting point \cite{Benzi1989}.   One natural question to
pose when there is an underlying particle description is how to derive
the above  dynamic equations from the  underlying microscopic dynamics
of the  system.  The Theory  of Coarse-Graining (ToCG), also  known as
Non-Equilibrium Statistical Mechanics,  or the Mori-Zwanzig formalism,
is  a well-established  framework  for the  derivation of  macroscopic
equations   from   the   underlying   microscopic   laws   of   motion
\cite{Grabert1982}.  This  theory allows one to  obtain closed dynamic
equations for  a set of  coarse variables  which are functions  of the
microscopic state of  the system.  From the point of  view of the ToCG
an equation like  (\ref{SPDE}) and, in general,  any statistical field
theory  in Soft  Matter  only makes  sense in  discrete  form and  the
partial differential equation \textit{appearance} should be taken as a
notational   convenience    \cite{Saarloos1982,Espanol1998a}.    Refs.
\cite{Zubarev1983,
  Espanol1998a,Espanol2009,Espanol2009c,DelaTorre2011}  have  explored
ways in which the program of the  ToCG can be implemented for the case
of  statistical  field  theories,  i.e.  how  the  equations  for  the
dynamics of stochastic  fields may be ``deduced''  from the underlying
microscopic dynamics of the  constituent particles.  Understanding the
microscopic  basis  of  a  mesoscopic equation  like  (\ref{SPDE})  is
essential  in order  to have  well-defined \textit{hybrid}  methods in
which  a  continuum-like  description   is  coupled  with  a  detailed
microscopic description.

One of  the important messages that we would  like to convey in
  the  present  paper is  that  there  is  a deep  connection  between
  ``numerical  analysis''  and  ``coarse-graining''  when  applied  to
  fluctuating  fluid systems.   Indeed,  from a  microscopic point  of
  view, the natural setting for a SPDE is a discrete one, where the CG
  hydrodynamic variables are  defined on the nodes of a  grid.  The CG
  variables are defined by assigning to every node the mass (momentum,
  energy) of the molecules that are "around" that node. There are many
  possibilities for this attribution.  The  simplest one is giving the
  mass of a molecule to the nearest node giving rise to a Voronoi cell
  partition of the molecules.  This  does not give physically sensible
  dynamic  equations as  we  have shown  in Ref.   \cite{Espanol2009}.
  Another  possibility  is to  use  a  finite element  basis  function
  defined on a triangulation  \cite{DelaTorre2011}.  While this solves
  the pitfalls of the Voronoi cell discretization, it corresponds to a
  lumped mass approximation of  the corresponding continuum equations.
  Here we  present a third  possibility that uses the  conjugate basis
  function   of  the   finite  element   basis  functions,   rendering
  microscopic  equations that  can  be  understood as  Petrov-Galerkin
  discretizations  of a  continuum  equation and,  therefore, have  an
  appropriate continuum limit. We discuss the microscopic
  basis  of  both  Eqs.    (\ref{PDE})  and  (\ref{SPDE})  within  the
  framework of  the Mori-Zwanzig  formalism that  provides microscopic
  expressions for all the objects in the equations and illustrates the
  different physical meaning of the  symbols in each equation.  After
having  formulated  the  discrete  version of  the  SPDE,  we  present
numerical simulations  in one dimension, based  on the Ginzburg-Landau
model for  the free energy  and discuss  the continuum limit  for this
model.

\section{Spatial discretization}
\label{sec:discPDE}

\subsection{Basis functions}
\label{sec:basis} As  a first step  for the discretization of  the PDE
(\ref{PDE}), we  introduce two basis  functions associated to  the two
operations involved in  the process of discretizing a  PDE.  The first
is  a discretization  operation  in which  a \textit{continuum  field}
$c({\bf r},t)$ is  reduced to a \textit{discrete field},  defined as a
set of discrete values ${\bf c}(t)=(c_1,\cdots,c_M)$, where each value
$c_\mu(t)$ is assigned to a position ${\bf r}_\mu$ of a node in a mesh
of $M$ points.  The second process is that of  transforming a discrete
field  into  a  continuum  field,  an  operation  also  understood  as
interpolation.  In the first process information is destroyed while in
the second it is created. These  two operations are implemented in the
present work through a set of two (dual or reciprocal) basis functions
\begin{align}  \boldsymbol{\delta}({\bf  r})&\equiv  \{\delta_\mu({\bf
r}),\mu=1,\cdots,  M\}  \nonumber\\ \boldsymbol{\psi}({\bf  r})&\equiv
\{\psi_\mu({\bf r}),\mu=1,\cdots, M\}
\end{align} The functions  $\psi_\mu({\bf r}),\delta_\mu({\bf r})$ are
localized around the node  point ${\bf r}_\mu$.  The discretization of
the continuum  field is given  by $ {\bf  c} =(\boldsymbol{\delta},c)$
where we introduce the scalar product as
\begin{align} (a,b)&=\int d{\bf r}a({\bf r})b({\bf r})
\end{align} whereas  the interpolation of the discrete  field is given
by the field  $ \overline{c}({\bf r})=\boldsymbol{\psi}({\bf r})\esc{\bf
c}$. In component form
we have
\begin{align}   c_\mu(t)&=\int  d{\bf  r}\delta_\mu({\bf   r})c({\bf  r},t)
\nonumber\\ \overline{c}({\bf r},t)&=\sum_\mu^M\psi_\mu({\bf r})c_\mu(t)
\label{natural}
\end{align}  A  natural  requirement  to   be  satisfied  by  the  two
operations of  discretizing and  interpolating is  that the  result of
interpolating  a discrete  field and  then discretizing  the resulting
interpolated field  should give  the original  discrete field.   It is
straightforward  to  show that  this  requirement  implies the  mutual
orthogonality of the basis functions
\begin{align} (\delta_\mu,\psi_\nu)&=\delta_{\mu\nu}
\label{projection0}
\end{align}  We  will  further  assume that  the  interpolation  basis
$\boldsymbol{\psi}({\bf r})$ is  \textit{linearly consistent}, meaning
\begin{align}
  \sum_\mu^M\psi_\mu({\bf r})&=1,&&  \int d{\bf r}\delta_\mu({\bf r})=1,
\nonumber\\
  \sum_\mu^M\psi_\mu({\bf r}){\bf r}_\mu
&={\bf r},&&\int d{\bf r}\,{\bf r}\delta_\mu({\bf r})={\bf r}_\mu.
\label{lc}
\end{align}

\begin{figure}[h!]
    \includegraphics{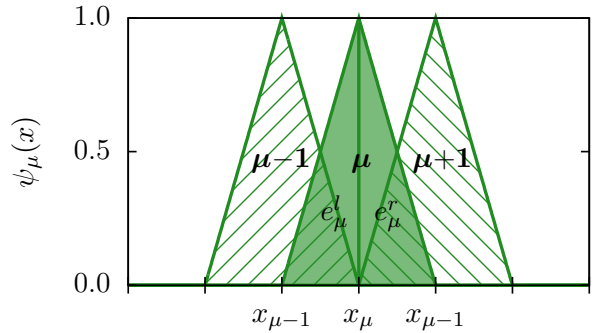}
    \caption{\label{fig:delaunay1D}The    linear    basis    functions
    $\psi_\mu(x)$ in 1D.  Each node $\mu$ has two elements,  $e_\mu^l$
    and $e_\mu^r$ shared with its neighbor nodes.}
\end{figure}

In  the present  paper, we  will  choose for  $\psi_\mu({\bf r})$  the
standard linear  basis function  of the finite  element on  node ${\bf
  r}_\mu$ that do satisfy the  linear consistency.  The finite element
is constructed from a triangulation of the grid like, for example, the
Delaunay   triangulation.    Fig.~{\ref{fig:delaunay1D}}   shows   the
functions  $\psi_\mu(x)$   for  three  neighbor  cells   in  1D.

\begin{figure}[h!]
    \includegraphics{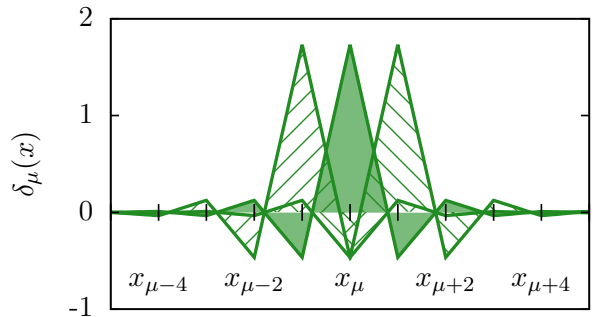} 
    \caption{\label{fig:delaunay1Ddelta} The conjugate basis functions
    $\delta_\mu(x)$  in 1D  in a  regular  lattice  with  total length
    $L=10$ and $M=64$ nodes.}
\end{figure}

We
further  assume that  the  basis functions  $\delta_\mu({\bf r})$  are
given  as linear  combinations of  the basis  functions $\psi_\mu({\bf
  r})$, this is
\begin{align}
  \delta_\mu({\bf r})&=M^\delta_{\mu\nu}\psi_\nu({\bf r})
\nonumber\\
\psi_\mu({\bf r})&=M^\psi_{\mu\nu}\delta_\nu({\bf r})
\label{lcdelta}
\end{align}
where the matrices  $M^\delta$ and $M^\psi$ are inverse  of each other
and, thanks to Eq. (\ref{projection0}), are given by
\begin{align}
  M^{\delta}_{\mu\nu}&=(\delta_\mu,\delta_\nu)
\nonumber\\
M^\psi_{\mu\nu}&=(\psi_\mu,\psi_\nu)
\label{Mdelta}
\end{align}
Fig.~{\ref{fig:delaunay1Ddelta}} shows the resulting functions $\delta_\mu(x)$.

In general, the  result of discretizing a field $c({\bf  r})$ and then
interpolating   this   discrete   field  gives   a   continuum   field
$\overline{c}({\bf  r})$ which  is different  from the  original field
$c({\bf  r})$,  except  for  some  particular  fields.   For  linearly
consistent basis functions, these  particular fields are linear fields
of  the form  $c({\bf r})=a  +{\bf  b}\cdot{\bf r}$  with $a,{\bf  b}$
constant.   Mathematically,   we  may  express  these   operations  as
convolutions
\begin{align}
  \overline{c}({\bf r})&=\int d{\bf r}'S({\bf r},{\bf r}') c({\bf r}')
\end{align}
where the smoothing kernel is defined as
\begin{align}
  S({\bf r},{\bf r}')&\equiv\sum_\mu^M\psi_\mu({\bf r})\delta_\mu({\bf r}')
=\sum_\mu^M\delta_\mu({\bf r})\psi_\mu({\bf r}')
\label{smoothing}
\end{align}
This smoothing operator satisfies
\begin{align}
  \int d{\bf r}' S({\bf r},{\bf r}')&=1
\nonumber\\
\int d{\bf r}'{\bf r}'S({\bf r},{\bf r}')&={\bf r}
\label{smoothprop}
\end{align}
and its effect on linear functions is much the same as the Dirac delta
function.  For  future reference, we  note that the first  equation in
(\ref{smoothprop})  can also  be  written as  the  partition of  unity
property
\begin{align}
  \sum_\mu{\cal V}_\mu\delta_\mu({\bf r})&=1
\label{voldelta}
\end{align}
where we have introduced the volume associated to node ${\bf r}_\mu$ as
\begin{align}
  {\cal V}_\mu&\equiv \int d{\bf r}\psi_\mu({\bf r})
\end{align}

\subsection{Petrov-Galerkin Weighted Residuals method}

A general  method for  discretizing partial differential  equations is
the Weighted Residual method \cite{Finlayson1972}. With the use of two
different  sets  of  basis  functions,  the method  is  known  as  the
Petrov-Galerkin  method.   The  idea   of  weighted  residuals  is  to
approximate the  actual solution  $c({\bf r},t)$ of  the PDE  with its
smoothed version $\overline{c}({\bf r},t)$, in such a way that
\begin{align}
  c({\bf r},t)&\approx\overline{c}({\bf r},t)
=\boldsymbol{\psi}({\bf r})\esc{\bf c}(t)
\label{closure}\end{align}
where now  ${\bf c}(t)=(\boldsymbol{\delta},c(\cdot,t))$ become  the unknown
of the  problem.  One defines the  residual of the PDE  (\ref{PDE}) as
the  result obtained  after  substitution in  Eq.  (\ref{PDE}) of  the
approximate field (\ref{closure})
\begin{align}
R({\bf r})&\equiv   \partial_t \overline{c}({\bf r},t) 
-\boldsymbol{\nabla}\esc \left[\Gamma(\overline{c}({\bf r},t))
 \boldsymbol{\nabla}\frac{\delta {\cal F}}{\delta c({\bf r})}[\overline{c}]\right]
\end{align}
By weighting the  residual with weights $\boldsymbol{\delta}({\bf r})$ 
and requiring the weighted residual to vanish we obtain
\begin{align}
\partial_t {\bf c}(t)& =
\left(\boldsymbol{\delta},\boldsymbol{\nabla}\esc\left[\Gamma(\overline{c}({\bf r},t))
\boldsymbol{\nabla}
\frac{\delta {\cal F}}{\delta c}[\overline{c}(t)]\right]\right)
\nonumber\\
&=
-\left(\boldsymbol{\nabla}\boldsymbol{\delta},\Gamma(\overline{c}({\bf r},t))
\boldsymbol{\nabla}
\frac{\delta {\cal F}}{\delta c}[\overline{c}(t)]\right)
\label{WRnol}\end{align}
where  an   integration  by  parts  has   been  performed.   Formally,
Eq. (\ref{WRnol}) is a set  of $M$ ordinary differential equations for
the $M$ unknowns ${\bf c}(t)$.

It is apparent  that we cannot proceed until we have  a way to compute
the functional derivative $\frac{\delta {\cal F}}{\delta c}$.  To this
end {\em define} the discrete free energy function $ F ({\bf c})$ as
\begin{align}
  F ({\bf c}) &\equiv {\cal F}\left[\boldsymbol{\psi}\esc{\bf c}\right]
\label{Fdef}
\end{align}
this is,  the free energy {\em  function} of the  discrete field ${\bf
  c}$ is  obtained by evaluating  the free energy {\em  functional} at
the interpolated field.  What we  need, though, is not  a discrete
approximation for  the functional,  but a  discrete approximation
for its functional  derivative. By using the functional  chain rule we
may compute the derivative of the function (\ref{Fdef})
\begin{align}
\frac{\partial  F }{\partial c_\mu}({\bf c}) &=
\int d{\bf r}'\frac{\delta  {\cal F}}{\delta c({\bf r}')}
\left[\boldsymbol{\psi}\esc{\bf c}\right]\psi_\mu({\bf r}')
\label{Fder}
\end{align}
 Let us  multiply
Eq. (\ref{Fder}) with  the basis function $\boldsymbol{\delta}({\bf r})$
\begin{align}
\boldsymbol{\delta}({\bf r})\esc  \frac{\partial F }{\partial {\bf c}}({\bf c}) &=
\int d{\bf r}'\frac{\delta  {\cal F}}{\delta c({\bf r}')}
\left[\boldsymbol{\psi}\esc{\bf c}\right]
S({\bf r},{\bf r}')
\label{funcder}
\end{align}
where  the  smoothing  kernel  $S({\bf r},{\bf  r}')$  is  defined  in
(\ref{smoothing}).  We will assume that the functional derivative does
not change appreciably within the range of $S({\bf r},{\bf r}')$.
In this  case, we  may
simply write from Eq. (\ref{funcder})
\begin{align}
\int d{\bf r}'\frac{\delta  {\cal F}}{\delta c({\bf r}')}
\left[\boldsymbol{\psi}\esc{\bf c}\right]
S({\bf r},{\bf r}')
&\approx
\frac{\delta  {\cal F}}{\delta c({\bf r})}
\left[\boldsymbol{\psi}\esc{\bf c}\right]
\label{funcderapp}
\end{align}
and, therefore, we have an approximate expression for the functional derivative
\begin{align}
 \frac{\delta {\cal F}}{\delta c({\bf r})}[\overline{c}]
=\frac{\delta {\cal F}}{\delta c({\bf r})}\left[\boldsymbol{\psi}\esc{\bf c}\right]
&\approx \boldsymbol{\delta}({\bf r})\esc  \frac{\partial F }{\partial {\bf c}}({\bf c})
\label{funcder2}
\end{align}
We  may  introduce  (\ref{funcder2})  into (\ref{WRnol}) and
obtain
\begin{align}
\frac{d{\bf c}}{dt}(t)
&=- {\bf D}({\bf c})\esc\frac{\partial F }{\partial {\bf c}}({\bf c})
\label{cfinOK}
\end{align}
where the dissipative matrix has the elements
\begin{align}
   D_{\mu\nu}({\bf c})&=\int d{\bf r}\boldsymbol{\nabla}\delta_{\mu}({\bf r}) \esc
\Gamma\left(\sum_\sigma\psi_\sigma({\bf r})c_\sigma\right)
\boldsymbol{\nabla}\delta_{\nu}({\bf r}) 
\label{Dfin}
\end{align}
The  matrix ${\bf  D}({\bf c})$  is manifestly  symmetric and
positive semi-definite because $\Gamma> 0$ (the semi character is due to (\ref{voldelta})).
The total number of particles, defined as $ N=\sum_\mu{\cal V}_\mu{\bf
  c}_\mu(t)$   is    a   dynamical    invariant   of    the   equation
(\ref{cfinOK}).  The  time  derivative  of the  discrete  free  energy
$F({\bf c}(t))$, which is given by
\begin{align}
  \frac{dF}{dt}({\bf c}(t))&=
-\frac{\partial F }{\partial {\bf c}}({\bf c})
\esc{\bf D}({\bf c})\esc\frac{\partial F }{\partial {\bf c}}({\bf c})\le 0
\end{align}
is  always   negative  or   zero,  because   ${\bf  D}({\bf   c})$  is
semi-positive   definite.   Therefore,   we  have   obtained  in   Eq.
(\ref{cfinOK}) a discrete version of the non-linear diffusion equation
(\ref{PDE})   that  captures   the   two   essential  features   about
conservation of the  number of particles and the Second Law.
As  we will  see,  the  fact that  the  diffusion  matrix is  positive
definite  in this  scheme is  crucial in  order to  construct discrete
versions of  SPDE.  Indeed, the fulfillment  of the Second Law  in the
form of an H-theorem in the  discrete setting is intrinsicly linked to
the possibility of describing thermal fluctuations.

\begin{figure}[t]
    \includegraphics{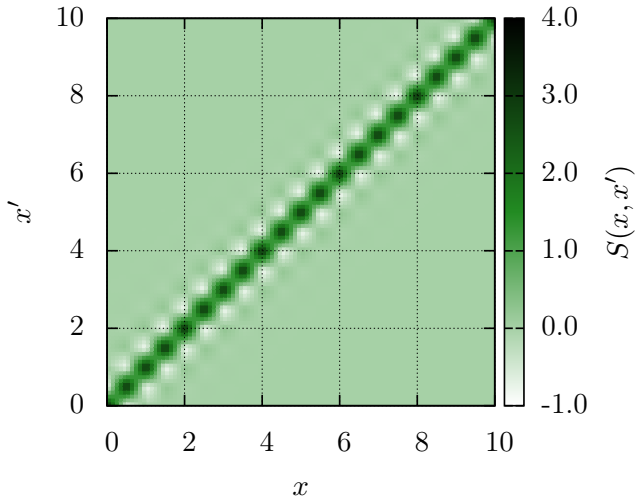}
    \caption{\label{figD}Plot of $S(x,x')$ in a 1D lattice with length
    $L=10$ and lattice  spacing $a = 0.5$.  The  function $S(x,x')$ is
    appreciably different from zero only for points differing by a few
    (2 or 3) lattice spacings.}
\end{figure}

The  only  approximation  that  we  have taken  is  that  the  functional
derivative of the  free energy functional hardly changes  in the range
of $S({\bf r},{\bf  r}')$ defined in Eq.   (\ref{smoothing}).  We have
plotted  this function  in Fig.   \ref{figD} and  observe that  if the
average  lattice spacing  is much  smaller  than the  length scale  of
variation of  the field, then the  approximation (\ref{funcder2}) will
be appropriate.  Of course, this  argument holds for the deterministic
setting where  the fields are smooth.  In a stochastic setting  as set
forth  later in  the  paper, for  which, in  general,  the fields  are
extremely  irregular the  procedure  should be  understood  not as  an
approximation but rather as a \textit{definition} of the discrete model itself.

\subsection{Explicit form of the dissipative matrix}
\begin{figure}[t]
      \rotatebox{0}{\resizebox{4cm}{!}{\includegraphics[width=2.5cm]{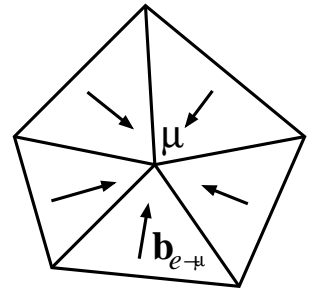}}} 
      \caption{In 2D, the Delaunay cell of node $\mu$ is surrounded by
      the triangular  elements $e$.  For each point  of the triangular
      element $e$,  there  is  a  constant  vector ${\bf b}_{e\to\mu}$
      that  points  towards   the  node  $\mu$  and   that  gives  the
      derivative of  the linear function  $\psi_\mu({\bf r})$  at that
      point.  }
 \label{Fig2}
\end{figure}

The  form (\ref{Dfin})  for the  dissipative matrix  involves  a space
integral that needs to be computed explicitly in order to introduce it
in a  computer code.  Note that  the mobility depends  on the position
through its dependence on the concentration field and, therefore, such
space  integrals  are  not  immediate.   We  will  use  the  following
approximation
\begin{align}
  \Gamma\left(\sum_\sigma \psi_\sigma({\bf r}) c_\sigma\right)&\approx
\sum_\sigma \psi_\sigma({\bf r})   \Gamma\left(c_\sigma\right) 
\label{approx}
\end{align}
in such a way that the  mobility function at the interpolated field is
approximated by a linear interpolation of the mobility function at the
nodes.   The  approximation is  exact  for  the points  ${\bf  r}={\bf
  r}_\mu$.  It is expected that  this approximation is appropriate for
smooth  functions   $\Gamma(c)$  provided   that  the  mesh   size  is
sufficiently small. With the  approximation in Eq. (\ref{approx}), the
dissipative matrix (\ref{Dfin}) becomes
\begin{align}
{\bf D}_{\mu\nu}({\bf c})&= \sum_{\sigma}  \Gamma(c_\sigma)\int d{\bf r}\boldsymbol{\nabla}\delta_{\mu}({\bf r}) \psi_\sigma({\bf r})
\boldsymbol{\nabla}\delta_{\nu}({\bf r})
\label{Ddeltamunu2}
\end{align}
The integral  is a geometric object  readily computable as  we show in
what  follows.  For  the   linear  finite  elements $\psi_\mu({\bf r})$,  we  may
explicitly compute the gradient of the basis functions
\begin{align}
\boldsymbol{\nabla}\delta_{\mu}({\bf r})&=M^\delta_{\mu\mu'}
\boldsymbol{\nabla}\psi_{\mu'}({\bf r})
\nonumber\\
\boldsymbol{\nabla}\psi_{\mu'}({\bf r})&=\sum_{e\in \mu'}  {\bf b}_{e\to\mu'}\theta_{e}({\bf r})
\label{nabladelta}
\end{align}
where  $\theta_{e}({\bf r})$  is  the characteristic  function of  the
sub-element $e$.   The gradient  of the basis  function $\psi_\mu({\bf
  r})$ is a constant vector ${\bf b}_{e\to\mu}$ for those points ${\bf
  r}$  that  are  within  the  sub-element  $e\in\mu$  of  node  $\mu$
\cite{Espanol2009f}.   In  Fig.    \ref{fig:delaunay1D}  we  show  the
sub-elements $e$ of the node $\mu$  in 1D while in Fig.  \ref{Fig2} we
show  the sub-elements  $e$ of  the node  $\mu$ and  the corresponding
vectors ${\bf b}_{e\to\mu}$ in 2D.

By using (\ref{nabladelta}) we have
\begin{align}
{\bf D}_{\mu\nu}({\bf c}) &= M^\delta_{\mu\mu'}M^\delta_{\nu\nu'}  
\sum_{e\in \mu',\nu'}  {\bf b}_{e\to\mu'}{\bf b}_{e\to\nu'}{\cal V}_e \Gamma_e({\bf c})
\label{Ddeltamunub}
\end{align}
where we have introduced the mobility $\Gamma_e$ of the element $e$ as
\begin{align}
  \Gamma_e({\bf c})&\equiv\sum_{\sigma\in e}W_{\sigma e} \Gamma(c_\sigma)
\label{mobe1}
\end{align}
and represents  a weighted average  of the mobility associated  to the
nodes $\sigma$  that are  the vertices  of the  element $e$.   We have
introduced  the  volume  of  element   $e$  and  the  geometric  ratio
$W_{e\sigma}$ as
\begin{align}
  {\cal V}_e&\equiv \int d{\bf r} \theta_{e}({\bf r})
\nonumber\\
   W_{\sigma e}&\equiv\frac{\int d{\bf r}
\theta_{e}({\bf r})\psi_\sigma({\bf r}) }{\int d{\bf r}
\theta_{e}({\bf r})}
\end{align}
In the  simulations presented in this  paper, we will assume  that the
mobility $\Gamma({\bf  c})=\frac{D c_0}{k_B  T}$ is a  constant, where
$D$ is a constant diffusion coefficient and $c_0$ is the equilibrium
value  of the  concentration  field.  In  this  case, the  dissipative
matrix (\ref{Ddeltamunu2}) is simply
\begin{align}
{\bf D}_{\mu\nu}({\bf c}) &= 
 \frac{D c_0}{k_B T}  \int d{\bf r} \boldsymbol{\nabla} \delta_{\mu}({\bf r})
\boldsymbol{\nabla} \delta_{\nu}({\bf r})= \frac{D c_0}{k_B T} {\bf L}_{\mu\nu}^{\delta} 
\label{eq:Dcte}
\end{align}
where the stiffness matrix ${\bf L}_{\mu\nu}^{\delta} $ is given by
\begin{align}
{\bf L}_{\mu\nu}^{\delta} &\equiv   \int d{\bf r} \boldsymbol{\nabla} \delta_{\mu}({\bf r})
\boldsymbol{\nabla} \delta_{\nu}({\bf r}).
\end{align}

\section{Physical interpretation from the Theory of Coarse-Graining}
\label{Sec:ToCG}
Up to now, the derivation  of the discrete equation (\ref{cfinOK}) out
of the PDE (\ref{PDE}) has been  an exercise in numerical analysis. In
this section,  we argue that a  microscopic view to the  problem shows
that there  is more physics  in (\ref{cfinOK}) than  this mathematical
operations suggest.   In the ToCG  the first  and crucial step  is the
selection  of the  (slow) CG  variables  while the  theory takes  care
automatically of the resulting dynamic  equations. We will see in this
section  that  the  above   numerical  analysis  suggests  \textit{the
  appropriate definition of the CG variables}  in the ToCG in order to
recover, from  microscopic grounds, a discrete  equation governing the
CG variables  identical to  (\ref{cfinOK}) which, by  construction, is
compatible with a continuum limit.

By using the  ToCG, the non-linear diffusion  equation (\ref{PDE}) can
be obtained  from microscopic  principles for  the description  of the
dynamics    of     a    colloidal    suspension,    as     shown    in
Ref.  \cite{Espanol2009c}.   One  chooses  as  relevant  variable  the
empirical or instantaneous concentration
\begin{align}
  \hat{c}_{\bf r}(z)&\equiv\sum_i^N\delta({\bf r}-{\bf r}_i)
\label{hatc}
\end{align}
where $z$  is the microscopic state  of the system and  ${\bf r}_i$ is
the  position of  the $i$-th  colloidal particle.   The ToCG  allows to
obtain an  exact equation for  the ensemble average $c({\bf  r},t)$ of
$\hat{c}_{\bf r}(z)$,  where the average  is over the solution  of the
microscopic  Liouville  equation.  The  resulting  exact  equation  is
non-local  in  space and  in  time.   Under  the assumption  that  the
concentration evolves very slowly as  compared with any other variable
in  the system,  the  exact integro-differential  equation becomes  an
approximate local in time equation.   A further approximation in which
the space non-locality of the  diffusion kernel is neglected, leads to
Eq.  (\ref{PDE})  \cite{Espanol2009c}.  The average $c({\bf  r},t)$ is
just the probability  density of finding (any)  one colloidal particle
(i.e.  its center of mass) at the  point ${\bf r}$ of space.  The free
energy functional ${\cal F}[c]$ and the mobility $\Gamma(c)$ have both
expressions in  terms of  the Hamiltonian  dynamics of  the underlying
system.  In particular, the mobility is given in terms of a Green-Kubo
expression (not shown  here) while ${\cal F}[c]$ is  the standard free
energy   density  functional   familiar  from   liquid  state   theory
\cite{Hansen1986}
\begin{align}
  {\cal F}[c]&=-k_BT\ln \int dz \rho^{\rm eq}(z)\exp\left\{-\int d{\bf r}\lambda({\bf r}) 
\hat{c}_{\bf r}(z)\right\}
\nonumber\\
&-\int d{\bf r}\lambda({\bf r})c({\bf r})
\label{DFT}
\end{align}
where $\rho^{\rm eq}(z)=\frac{1}{Z}e^{-\beta H(z)}$ is the equilibrium
canonical ensemble of the system with Hamiltonian $H(z)$, $z$ denoting
the microstate,  i.e. positions and  momenta of  all the atoms  of the
system. The  Lagrange multiplier  $\lambda({\bf r})$  is fixed  by the
condition
\begin{align}
  \frac{\delta {\cal F}}{\delta c({\bf r})}[c]&=\lambda({\bf r})
\end{align}
that  connects  in  a  one  to  one  manner  the  Lagrange  multiplier
$\lambda({\bf r})$, usually referred to as the chemical potential, and
the  average value  $c({\bf r)}$  of (\ref{hatc}).     Note
that  Eq. (\ref{PDE})  describes  a dynamics  in  which ${\cal  F}[c]$
always decreases,  while keeping the  number of particles  fixed.  The
final  equilibrium  profile  is  then  obtained  after  a  constrained
maximization of  ${\cal F}[c]$  with the  result that  the equilibrium
concentration field is such  that the chemical potential $\lambda({\bf
  r})$ is constant in space.
\color{black}

Instead of  using the relevant  variables (\ref{hatc}) we may  use the
number of colloidal  particles per unit volume at the  mesh node ${\bf
  r}_\mu$ as input for the ToCG. These relevant variables are
\begin{align}
  \hat{c}_\mu(z)&\equiv\sum_i^N\delta_{\mu}({\bf r}_i)
\label{hatcmu}
\end{align}
The function $\delta_{\mu}({\bf r})$ is assumed to be localized around
the mesh node ${\bf r}_\mu$ and, therefore, the function $\hat{c}_\mu$
counts  the  number  of  colloidal particles  that  are  around  ${\bf
  r}_\mu$.   Different   functional  form   have  been   proposed  for
$\delta_\mu({\bf r})$, ranging  from a function defined in  terms of a
finite   number   of   Fourier   modes   \cite{Zubarev1983}   to   the
characteristic  function  of the  Voronoi  cell  around ${\bf  r}_\mu$
(divided by  the volume of  the cell) \cite{Saarloos1982}. As  we have
discussed in Ref.  \cite{Espanol2009}, the resulting dynamic equations
are not well behaved for the Voronoi cells.  Motivated by the previous
section, in  the present paper  we choose $\delta_\mu({\bf r})$  to be
the  basis  function  dual  to   the  finite  element  basis  function
$\psi_\mu({\bf  r})$. The  proposal for  $\delta_\mu({\bf r})$  in the
present paper  differs from our previous  proposal \cite{Espanol2009c}
where  we used  $\delta_\mu({\bf r})=\psi_\mu({\bf  r})/{\cal V}_\mu$.
The former selection  is equivalent to a lumped  mass approximation in
which the  mass matrix is  approximated by  a diagonal matrix,  and is
reasonable for regular lattices.  A lumped mass approximation does not
satisfy  the  natural  requirement  leading to  the  orthogonality  in
Eq.  (\ref{projection0}) and  for this  reason  we will  use the  dual
$\delta_\mu({\bf  r})$ given  in (\ref{lcdelta}).   As the  resolution
increases and the  number of node points ${\bf  r}_\mu$ increases, the
support  of the  function $\delta_\mu({\bf  r})$ is  reduced and  this
function converges  weakly to the  Dirac delta function.  In  the high
resolution  limit ($M\to\infty$) the  function (\ref{hatcmu})  converges  weakly  to
(\ref{hatc}).   Note  that  due   to  (\ref{voldelta}),  the  relevant
variables (\ref{hatcmu}) satisfy
\begin{align}
  \sum_\mu^M{\cal V}_\mu \hat{c}_\mu(z)&=N
\label{forall}
\end{align}
irrespective  of  the value  of  the  microstate $z$.   The  variables
$\hat{c}_\mu$  change stochastically  as  a result  of the  stochastic
motion of the underlying colloidal particles.

In the next two subsections, we  present the results obtained from the
ToCG regarding the dynamics for the averages ${\bf c}$ of the discrete
variables  (\ref{hatcmu}) and  for the  distribution function  $P({\bf
  c},t)$ of  these variables   (or the  corresponding SDE).
These dynamic  equations are (\ref{CGcdot})  for the average  value of
the  discrete ${\bf  c}$  and (\ref{eq:SDEdisc})  for the  fluctuating
variables, and are \color{black}  the discrete versions of (\ref{PDE})
and (\ref{SPDE}), respectively. 

\subsection{Physical interpretation of the PDE}
 The  ToCG
allows us to obtain closed  equations of motion for the time dependent
\textit{ensemble  average} ${\bf c}(t)=\langle  \hat{\bf c}(t)\rangle$
of the  discrete variables (\ref{hatcmu}).  In this  case, one obtains
\begin{align}
  \frac{d{\bf c}}{dt}(t)&=-\overline{\bf D}({\bf c})
\frac{\partial \overline{F}({\bf c})}{\partial {\bf c}}
\label{CGcdot}
\end{align}
The \textit{renormalized}  diffusion matrix $\overline{\bf  D}({\bf c})$ is
given by  a Green-Kubo formula (not shown  here). The \textit{renormalized}
free energy function $\overline{F}({\bf c})$ is defined as
\begin{align}
\overline{F}({\bf c}) &=-k_BT\ln \int dz \rho^{\rm eq}(z)\exp\{-\boldsymbol{\lambda}\esc\hat{\bf c}(z)\}-
\boldsymbol{\lambda}\esc{\bf c}
\label{Fdress}
\end{align}
The conjugate parameters $\boldsymbol{\lambda}$ are
in one to one connection with the discrete field ${\bf c}$ through the relation
\begin{align}
  \frac{\partial \overline{F}}{\partial {\bf c}}({\bf c})&=\boldsymbol{\lambda}
\end{align}

The equation of motion (\ref{CGcdot}) obtained microscopically has the
same structure  as the discretized Eq.  (\ref{cfinOK}).   In fact, the
symbol  ${\bf c}(t)$  in these  two  equations has  the same  physical
meaning because the relevant variables (\ref{hatc}) and (\ref{hatcmu})
are related linearly
\begin{align}
  \hat{c}_\mu(z)&=\int d{\bf r}\delta_\mu({\bf r})\hat{c}_{\bf r}(z)
\label{cc}
\end{align}
Therefore, a natural  question is: What is the  connection between the
discrete  free energy  $F({\bf c})$  defined ``numerically''  from the
free energy functional ${\cal F}[c]$ through Eq.  (\ref{Fdef}) and the
renormalized  free  energy  function $\overline{F}({\bf  c})$  defined
``physically''  in  Eq.  (\ref{Fdress})?   In  the  remaining of  this
section we  show that $\overline{F}({\bf c})\mapsto F({\bf c})$  in the limit
of high resolution.

The idea  is as follows.  We  have two levels of  description, Level 1
given in  terms of the  relevant variables $\hat{c}_{\bf r}(z)$  and a
more coarse-grained Level  2 given in terms of  the relevant variables
$\hat{c}_\mu(z)$. Because  the relevant variables of  these two levels
of description are related linearly,  i.e. Eq.  (\ref{cc}), the bridge
theorem \cite{Anero2013},  also known as the  contraction principle in
large-deviation  theory  \cite{Touchette2009},  applies and  the  free
energy of  Level 2  is obtained  from that of  Level 1,  by maximizing
${\cal  F}[c]$  subject  to a  given ${\bf  c}$.   We  extremize  without
restriction
\begin{align}
  {\cal F}[c]-\boldsymbol{\lambda}\esc
\int d{\bf r}\boldsymbol{\delta}({\bf r}) c({\bf r})
+\mu\int d{\bf r}c({\bf r})
\end{align}
The maximizer  $c^*({\bf r})$ of  this functional depends on the Lagrange
multipliers  $\boldsymbol{\lambda},\mu$ which  are fixed  by requiring
the constraints
\begin{align}
  \int d{\bf r}\boldsymbol{\delta}({\bf r}) c({\bf r})&={\bf c}
\nonumber\\
\int d{\bf r}c({\bf r})&=1
\label{norm}
\end{align}
Therefore, the maximizer $c^*({\bf c})$ depends implicitly on ${\bf c}$.
The bridge theorem ensures that the free energy of Level 2 is obtained
when we evaluate the free energy of Level 1 at $c^*({\bf c})$
\begin{align}
\overline{F}({\bf c})&={\cal F}[c^*({\bf c})]
\label{exact}
\end{align}
This  is  an   exact  result.   We   now  show  that  $
\overline{F}({\bf c})\approx F({\bf  c})$ as the resolution increases.
To this end, we write
\begin{align}
  c^*({\bf r})&= \boldsymbol{\psi}({\bf r}) \esc{\bf c}+{\cal \epsilon}({\bf r}) 
\label{c*}
\end{align}
where therefore, the error field  ${\cal \epsilon}({\bf r})$ defined through this
equation  is  the  difference  between  the  solution  $c^*$  and  the
interpolated  field.  By inserting  (\ref{c*})  into (\ref{exact})  we
obtain
\begin{align}
  F({\bf c})&={\cal F}[c^*]={\cal F}[\boldsymbol{\psi}\esc{\bf c}+{\cal \epsilon}]
\nonumber\\
&={\cal F}[\boldsymbol{\psi}\esc{\bf n}]+\int d{\bf r}{\cal \epsilon}({\bf r})
\frac{\delta {\cal F}}{\delta c({\bf r})}[\boldsymbol{\psi}\esc{\bf c}]+\cdots
\nonumber\\
&= \overline{F}({\bf c})+\int d{\bf r}{\cal \epsilon}({\bf r})
\frac{\delta {\cal F}}{\delta c({\bf r})}[\boldsymbol{\psi}\esc{\bf c}]+\cdots
\label{exact2}
\end{align}
Now we show that ${\cal  \epsilon}({\bf r})$ is vanishingly  small in
the  high resolution  limit. 
The  solution field $c^*({\bf r})$ has to  fulfill the
restriction (\ref{norm}) and, therefore, we have
\begin{align}
{\bf c}&= \int d{\bf r}\boldsymbol{\delta}({\bf r})c^*({\bf c})({\bf r})
\nonumber\\
&=\int d{\bf r}\boldsymbol{\delta}({\bf r}) \left(\boldsymbol{\psi}({\bf r})\esc{\bf c}\right)+
\int d{\bf r}\boldsymbol{\delta}({\bf r}){\cal \epsilon}({\bf r})
\end{align}
The orthogonality of the  basis functions (\ref{projection0})  implies that 
\begin{align}
  \int d{\bf r}\delta_\mu({\bf r}){\cal \epsilon}({\bf r})=0\quad \forall \mu
\end{align}
and the error field converges weakly to zero.
As  we increase  the resolution,  the functions  $\delta_\mu({\bf r})$
become  more and  more localized,  implying that  the function  ${\cal
  \epsilon}({\bf  r})$ is  vanishingly  small in  the high  resolution
limit.  In this limit, therefore,  the renormalized free energy in the
CG method $\overline{F}({\bf c})$ can be obtained in a simpler way not
through the  exact Eq. (\ref{exact}),  but rather through  the simpler
recipe (\ref{Fdef}).  This  is very convenient because  there are many
good  approximate  free  energy   density  functionals  ${\cal  F}[c]$
available in the literature.

   Note  that  the  above argument  applies  for  arbitrary
functionals, in general non-local.  Non-locality arises usually in 
models for free  energy functionals due to the  appearance of smoothed
concentration fields that involve integrals of the concentration field
with  weight  functions \cite{Hansen1986}.   As  we  show in  Appendix
\ref{ap:approx-nonlinear} for  the case  of the  exact 1D  Percus free
energy functional for  hard rods, there is no problem  in dealing with
these      non-local       functionals      with       the      recipe
(\ref{Fdef}). Nevertheless, in the present  paper we will not consider
these non-local functionals since considering non-locality only in the
free-energy functional but not in the mobility operator is not physically
consistent. At the same time, a detailed microscopic understanding of non-local
mobility operators (and thus non-local noise correlations) is lacking
and requires a careful study in the future.  \color{black}

\subsection{Physical interpretation of the SPDE}
We now justify the SPDE (\ref{SPDE}) from a physical perspective along
the lines in  the previous section.  The first question  to address is
the physical  meaning to be assigned  to the symbol $c({\bf  r},t)$ in
Eq.  (\ref{SPDE}).  It cannot be  ``the probability density of finding
a  colloidal  particle   at  ${\bf  r}$  at  time  $t$''   as  in  Eq.
(\ref{PDE}), because in (\ref{SPDE})  $c({\bf r},t)$ is an intrinsicly
stochastic field and cannot be  a ``fluctuating probability''.  Except for
non-interacting   Brownian  walkers,   Eq.   (\ref{SPDE})   cannot  be
understood as  an equation governing  the dynamics of the  spiky field
(\ref{hatc})  and  even  in  this   case,  (\ref{SPDE})  can  only  be
interpreted formally \cite{Donev2014}. There has  been a lot of debate
about  the meaning  of  fluctuating  equations in  the  field of  DDFT
\cite{Archer2004a}.

Clearly,  in order  to speak  about  ``fluctuations in  the number  of
particles  per   unit  volume''  one   needs  to  use   the  variables
(\ref{hatcmu}) as  relevant variables and consider  the time dependent
probability  distribution  $P({\bf  c},t)$ that  the  phase  functions
$\hat{\bf c}_\mu(z)$  in (\ref{hatcmu})  take particular  values ${\bf
  c}$.    From  the   ToCG  it   is  possible   to  obtain   an  exact
integro-differential   equation  for   $P({\bf   c},t)$.   After   the
assumption of clear separation of time scales between the evolution of
the concentration  and any other  variable in the system,  one obtains
the following Fokker-Planck equation that governs $P({\bf c},t)$
\begin{align}
  \partial_tP({\bf c},t) &=
\frac{\partial}{\partial {\bf c}}\esc\left\{\hat{\bf D}({\bf c})\esc
\left[\frac{\partial \hat{F} }{\partial {\bf c}}({\bf c})P({\bf c},t)
+k_BT
\frac{\partial}{\partial {\bf c}}P({\bf c},t)\right]\right\}
\label{eq:FPE}
\end{align}
The {\em bare}  diffusion matrix $\hat{\bf D}({\bf c})$  is defined in
terms   of  a   Green-Kubo  expression   (not  shown)   and  satisfies
$\sum_\mu{\cal  V}_\mu\hat{\bf  D}_{\mu\nu}({\bf  c})=0$ where  ${\cal
  V}_\mu$ is  the volume associated to cell  $\mu$.  
The \textit{bare}
$\hat{\bf D}({\bf  c})$ is,  in  general, a  quantity  different from  the
renormalized  $\overline{\bf D}({\bf  c})$  in Eq.   (\ref{CGcdot}).   

The
\textit{bare}  free  energy $\hat{F}({\bf  c})$  is  defined from  the
equilibrium distribution of (\ref{eq:FPE})
\begin{align}
  P^{\rm eq}({\bf c}) 
&={ \frac{1}{\hat{Z}}}\delta\left(\sum_\mu{\cal V}_\mu c_\mu-N\right)
\exp\left\{-\frac{1}{k_BT}\hat{F}({\bf c})\right\}
\label{Einsdisc}
\end{align}
{ where  $\hat{Z}$ is  the normalization.}   The Dirac  delta contribution
reflects  the mass  conservation (\ref{forall})  and ensures  that the
probability vanishes for  those concentration fields that  do not have
exactly  $N$ particles.   Note that  microscopically, the  equilibrium
distribution function is given by the phase space integral
\begin{align}
  P^{\rm eq}({\bf c})   &=\int dz\rho^{\rm eq}(z)\delta({\bf c}-\hat{\bf c}(z))
\label{Peq1}
\end{align}
The bare free energy $\hat{F}({\bf  c})$ is, in general, a function of
${\bf   c}$  which  is   different  from   the renormalized   free  energy
$\overline{F}({\bf c})$ { since its microscopic definition is different (see additional discussion below).
The difference between the
two is expected to be larger the larger the fluctuations are (i.e., the smaller the coarse-graining cells are).}

The  Ito stochastic
differential equation (SDE) corresponding to the FPE (\ref{eq:FPE}) is 
\begin{align}
d{\bf c}(t)
&=- \hat{\bf D}({\bf c})\esc\frac{\partial \hat{F} }{\partial {\bf c}}({\bf c})dt 
+k_BT\frac{\partial }{\partial {\bf c}}\esc\hat{\bf D} ({\bf c})dt+d\tilde{\bf c}(t)
\label{eq:SDEdisc}
\end{align}
where  the term  proportional to  $k_BT$ is  a reflection  of the  Ito
stochastic interpretation of  this SDE.  Here $d{\tilde{\bf  c}}$ is a
linear  combination  of  Wiener  processes  that  has  the  covariance
structure
\begin{align}
\left\langle  \frac{d\tilde{\bf c}}{dt}(t)  \frac{d\tilde{\bf c}}{dt}(t')\right\rangle &=2k_BT\hat{\bf D}({\bf c})\delta(t-t')
\label{FD}
\end{align}

\subsection{Discussion}
   We  have  obtained,  from   the  ToCG  the  two  dynamic
equations, (\ref{CGcdot})  for the  average of the  discrete variables
(\ref{hatcmu}), and (\ref{eq:SDEdisc}) for the fluctuating dynamics of
these discrete variables. The  structure of (\ref{CGcdot}) is formally
identical to  the finite  element discretization of  (\ref{PDE}). This
suggests  that  the continuum  limit  of  the microscopically  derived
(\ref{CGcdot}) is well-defined and given by (\ref{PDE}).  In this way,
the definition (\ref{hatcmu}) of the CG variables with the appropriate
basis functions is crucial in  order to obtain discrete equations with
proper continuum limit.  The discussion of the continuum  limit of the
stochastic equation (\ref{eq:SDEdisc}) is more subtle and given below.

\color{black}
The  ToCG  is extremely  useful  as  it  gives  the structure  of  the
equations (\ref{PDE}), (\ref{CGcdot}),  and (\ref{eq:SDEdisc}), but it
remains  formal because  the microscopic  expressions for  the objects
appearing    in    these     equations    ${\cal    F}[c],\hat{F}({\bf
  c}),\overline{F}({\bf    c})$   and    $\Gamma(c),\hat{\bf   D}({\bf
  c}),\overline{\bf  D}({\bf  c})$ are  too  complex  to be  evaluated
explicitly.   Some  general features  may  be  inferred, though.   For
example,  there  exists  an  exact connection  between  the  bare  and
renormalized free  energies, which  can be  obtained by  inserting the
identity $\int  d{\bf c}\delta({\bf  c}-\hat{\bf c}(z))=1$  inside Eq.
(\ref{Fdress})  and  using  Eqs. (\ref{Peq1}),  (\ref{Einsdisc}).  The
result is
\begin{align}
e^{-\beta \overline{F}({\bf c})-\boldsymbol{\lambda}({\bf c})\esc{\bf c}}&=\int d{\bf c}'\delta(\boldsymbol{{\cal V}}\esc{\bf c}'-N)
{ \frac{1}{\hat{Z}}}e^{-\beta\hat{F}({\bf c}')-\boldsymbol{\lambda}({\bf c})\esc{\bf c}'}
\label{FhatF}
\end{align}

Both the  renormalized $\overline{F}({\bf c})$ and  bare $\hat{F}({\bf
  c})$ free  energies depend in  a non-trivial  way on the  cell size,
introduced implicitly through $\delta_\mu({\bf  r})$ in the definition
of the CG variable ${\bf c}$ in Eq.  (\ref{hatcmu}).  We have shown in
the   previous    section   that   the   renormalized    free   energy
$\overline{F}({\bf c})$ may be obtained  from the standard free energy
functional  ${\cal F}[c]$  according  to $\overline{F}({\bf  c})={\cal
  F}[\boldsymbol{\psi} \esc  {\bf c}]$, in the  high resolution limit.
It is, therefore,  legitimate to ask whether there exists  a bare free
energy \textit{functional}  $\hat{\cal F}[c]$ such that  the bare free
energy \textit{function} may be also written in the form $\hat{F}({\bf
  c})=\hat{\cal F}[\boldsymbol{\psi}  \esc {\bf  c}]$.  Unfortunately,
the answer is  not straightforward.  For example, looking  at the high
resolution limit of (\ref{FhatF}) makes  not much sense because in the
continuum  limit the  variables  (\ref{hatcmu}) become  spiky like  in
(\ref{hatc})  and the  Dirac delta  function $\delta({\bf  c}-\hat{\bf
  c}(z))$ gives a  probability $P^{\rm eq}({\bf c})$  that is non-zero
only  for  spiky  fields,  parametrized   with  the  position  of  the
particles.   In  fact,  this  probability is  given,  up  to  particle
permutations,   by  the   Gibbs  ensemble   $e^{-\beta  H(z)}$.    For
non-interacting particles, in this limit (\ref{SPDE}) becomes simply a
formal rewriting of the underlying particle dynamics \cite{Donev2014}.
In that  limit (\ref{FhatF}) falls  back to (\ref{Fdress}) which  is a
trivial result.  In addition, the separation of time scales underlying
the Markov approximation  and the FPE is expected to  fail in the high
resolution limit.

On the other  hand, in lower resolution situations, when  we expect to
have typically many particles per cell and the equilibrium probability
(\ref{Peq1}) should remain highly peaked,  we may compute the integral
in (\ref{FhatF}) with a saddle approximation, giving
\begin{align}
\hat{F}({\bf c})\approx   \overline{F}({\bf c})
\end{align} { up to an  irrelevant constant}. This equation would allow  one to find the functional form  of the bare free
energy  $\hat{F}({\bf  c})$  should  the renormalized  free  energy  $
\overline{F}({\bf  c})$  be known.   Of  course,  for low  resolutions
(large cell volumes), we do not know in general the functional form of
the renormalized free energy. Only for single phase systems with known
macroscopic  thermodynamics one  may  use local  models  for the  free
energy  functional in  terms of  the macroscopic  free energy  density
$f^{\rm eq}(c)$ of the form
\begin{align}
 \hat{F}({\bf c})&=\int d{\bf r}f^{\rm eq}(\psi_\mu({\bf r})c_\mu)
\end{align}

This leaves us in the position of
having to  \textit{model} the bare free energy  $\hat{F}({\bf c})$. In
the present paper, we will model the bare free energy according to the
prescription
\begin{align}
  \hat{F}({\bf c})&=\hat{\cal F}[\boldsymbol{\psi}\esc{\bf c}]
\label{bareFdef}
\end{align}
for  a  supposedly  known   bare  free  energy  functional  $\hat{\cal
  F}[c]$. With this,  we are assuming that  all ``resolution dependent
features''   of  the   probability  $P^{\rm   eq}({\bf  c})$   in  Eq.
(\ref{Peq1}) can be dealt with  a single functional $\hat{\cal F}[c]$.
Whether  the actual  probability $P^{\rm  eq}({\bf c})$  is accurately
given by  such a model  is a completely open  question that we  do not
address         in        the         present        work         (see
\cite{Troster2007a},\cite{Troster2007}).   However,  this  is  current
practice in  the literature  starting from,  for example,  the seminal
work by van Kampen's on the calculation of $P^{\rm eq}({\bf c})$ for a
van der Waals fluid \cite{Kampen1964,Reichl1980,Espanol2001}.  He used
for  $\delta_\mu({\bf r})$  the  Voronoi  characteristic function  and
obtained an approximate  expression for $P^{\rm eq}({\bf  c})$ that is
written without much explanation in a continuum form (see Eq.  (13) of
Ref. \cite{Kampen1964}).  Such a  happy transition from discrete world
to continuum notation  is usual \cite{Saarloos1982} but  not exempt of
potential problems. The natural question  to ask is whether a sequence
of equations like (\ref{eq:SDEdisc}), with  a given model for the bare
free energy functional $\hat{\cal F}[c]$  has a ``continuum limit'' as
we increase the resolution, in such a way that a proper meaning can be
given  to an  equation like  (\ref{SPDE}).  Even  if such  a continuum
limit is  obtained, we should expect  that, in general, the  bare free
energy functional $\hat{\cal F}[c]$ will not  be the same as the usual
free  energy  functional  ${\cal   F}[c]$  in  Eq.   (\ref{DFT}).   In
particular, note  from (\ref{FhatF})  that ${\cal  F}[c]$ is  always a
convex  functional  even though  $\hat{\cal  F}[c]$  needs not  to  be
convex.

\section{Thermal fluctuations}

  The  bare and  renormalized diffusion  matrices $\hat{\bf
  D}_{\mu\nu}$,  $\overline{\bf D}_{\mu\nu}$  are  given  in terms  of
Green-Kubo expressions and  depend in general on the  state ${\bf c}$.
They are, in principle, different quantities. However, for the sake of
simplicity, in  the present  paper we assume  that the  bare diffusion
matrix  $\hat{\bf   D}_{\mu\nu}$  has   the  same  structure   as  the
renormalized  diffusion  matrix  $\overline{\bf  D}_{\mu\nu}$  in  Eq.
(\ref{Ddeltamunub}).   Furthermore, in  the simulation  results to  be
presented below, a constant mobility will be assumed.

\color{black}

The actual  form of the  diffusion matrix  determines the form  of the
noise terms through Eq.  (\ref{FD}).  The problem that we solve in the
present section  is how  to compute  explicitly the  stochastic forces
$d\tilde{\bf  c}_\mu$  satisfying (\ref{FD}).   We  need  to find  the
particular linear  combination of  Wiener processes  that lead  to Eq.
(\ref{FD}).  A  brute-force calculation of  the square root  matrix of
the  diffusion  matrix  ${\bf  D}$  is  very  costly  computationally,
especially if it  depends on the state ${\bf c}$ \cite{Plunkett}.  Instead, we propose
an explicit formula inspired by the  very structure of the random term
(\ref{J}) in  the continuum equation.   Alternatively, we may  look at
the explicit  structure of  the so called  \textit{projected currents}
that  enter the  Green-Kubo  expression  \cite{Espanol2009c}. In  both
cases, we obtain the same result that we detail below.

Recall that in  the Weighted Residual procedure we  multiplied the PDE
with $\delta_\mu({\bf r})$  and integrated over space.  If  we do this
for  the  stochastic  term $\boldsymbol{\nabla}\esc\tilde{\bf  J}$  in
Eq. (\ref{J}) we obtain
\begin{align}
  \frac{d{\tilde{\bf c}}_\mu  }{dt}&=-\int d{\bf r}\boldsymbol{\zeta}({\bf r},t)\esc\boldsymbol{\nabla}\delta_{\mu}({\bf r})
\sqrt{2 k_B T \Gamma(\overline{c}({\bf r},t))}
\label{czeta}
\end{align}
The correlations of the noises
(\ref{czeta})   are  easily   computed  under   the   assumption  that
$\boldsymbol{\zeta}({\bf r},t)$  is a white  noise in space  and time,
satisfying
\begin{align}
\langle  \boldsymbol{\zeta}({\bf r},t)\boldsymbol{\zeta}({\bf r}',t')\rangle &=\delta({\bf r}-{\bf r}')\delta(t-t')
\end{align}
The result is 
\begin{align}
\left\langle  
\frac{d{\tilde{\bf c}}_\mu}{dt}(t)  
\frac{d{\tilde{\bf c}}_\nu}{dt}(t')\right\rangle 
&=2k_BT \overline{\bf D}_{\mu\nu}({\bf c})\delta(t-t')
\end{align}
and, therefore, (\ref{czeta}) has  the desired covariances (\ref{FD}).
However      (\ref{czeta})      involves     the      white      noise
$\boldsymbol{\zeta}({\bf r},t)$  and an integral over  the whole space
while what  we are  looking for  is a linear  combination of  a finite
number  of   independent  Wiener  processes.   By   using  the  result
(\ref{nabladelta}) in (\ref{czeta}), and taking the same approximation
for   the  mobility   that   lead  to   (\ref{mobe1})   leads  us   to
\textit{postulate} the following linear combination of white noises
\begin{align}
\frac{d\tilde{\bf c}_\mu}{dt} &=\sum_\nu{\bf M}^\delta_{\mu\nu}\sum_{e\in\nu}
\sqrt{2 k_B T {\cal V}_e \Gamma_e({\bf c})}
{\bf b}_{e\to\nu}  \zeta_e(t)
\label{eq:finalnoise}
\end{align}
Here $\zeta_e(t)$ is an independent white noise
 associated to the element $e$ in the
triangulation, satisfying 
\begin{align}
\langle \zeta_e(t)\zeta_{e'}(t')  \rangle&=\delta_{ee'}\delta(t-t')
\end{align}
 It is a simple exercise to show that the covariance of the noises
(\ref{eq:finalnoise}) satisfies Eq.
(\ref{FD}).
The random term (\ref{eq:finalnoise}) respects mass conservation
in the sense that
\begin{align}
  \sum_\mu{\cal V}_\mu \frac{d\tilde{\bf c}_\mu}{dt}&=0
\end{align}
To   prove  this,   one   needs  to   use  $\sum_\mu{\cal   V}_\mu{\bf
  M}^\delta_{\mu\nu}=1$,  which   is  obtained  from   the  definition
(\ref{lcdelta}) and the  property (\ref{voldelta}), and $ \sum_{\mu\in
  e}{\bf b}_{e\to\mu}=0$.

\section{Functional form for the bare free energy function $\hat{F}({\bf c})$}
Colloidal suspensions  that may phase separate  in liquid-vapor phases
\cite{Tata1992}  may be  described  by  a van  der  Waals free  energy
functional. Near the critical point, the van der Waals free energy may
be  approximated by  a  Ginzburg-Landau model,  as  shown in  Appendix
\ref{ap:free}.  The bare free energy  functional that we will consider
in the present paper is the Ginzburg-Landau (GL) free energy
\begin{align}
  \hat{\cal F}^{{\rm (GL)}}[c({\bf r})] &=  k_B T\int
d{\bf r}\, \left\lbrace
\frac{r_0}{2}\phi({\bf r})^2 
+\frac{K}{2}  \left(\nabla \phi({\bf r})\right)^2\right\rbrace
\nonumber\\
&+k_B T\int
d{\bf r}\,  \frac{u_0}{4} \phi({\bf r})^4 
\label{eq:GLFEfunctional}
\end{align}
where  $\phi({\bf  r})  =  (c({\bf  r})-c_0)/c_0$  and  $c_0$  is  the
equilibrium concentration.  The reason to  use the GL model instead of
the original van  der Waals model in the present  work arises from our
interest in the  numerical aspects of the problem.   The GL \textit{polynomial}
model allows one to compute the bare free energy function $F({\bf c})$
exactly,    without     further    approximations.      In    Appendix
\ref{ap:approx-nonlinear}  we   discuss  possible   approximations  to
non-polynomial free energy functionals. The parameters in the GL model
in terms of the van der Waals model are (see Appendix \ref{ap:free})
\begin{align}
u_0&=\frac{3}{16b}
\nonumber\\
r_0 &=\frac{3}{4b}  \left(1-\frac{T_c}{T} \right)
\nonumber\\
K&= \frac{3}{4b}\sigma^2
\frac{T_c}{T} 
\label{param}
\end{align}
These  coefficients  depend  on  temperature  but are  assumed  to  be
independent  of the concentration  field. Here,  $b$ is  the molecular
volume of  the van  der Waals model,  $T_c$ the  critical temperature,
$\sigma$ is a length scale related to the range of the attractive part
of the microscopic potential.  

The GL free  energy functional is non-linear due to  the $\phi^4$ term
and  observables  like  correlation  functions can  only  be  computed
explicitly in  an approximate  way, either  by perturbation  theory or
other means. Two models that we will also consider in the present work
are the  solvable Gaussian  model with surface  tension (GA+$\sigma$),
which      is     obtained      by     setting      $u_0=0$     in
Eq. (\ref{eq:GLFEfunctional}), and the  Gaussian model without surface
tension (GA) which is obtained after setting $u_0=0,K=0$. They are
\begin{align}
  \hat{\cal F}^{(\mathrm{GA}+\sigma)}[ c({\bf r})] &= 
k_B T\int d{\bf r}\, \left\lbrace 
\frac{r_0}{2}\phi({\bf r})^2 
+\frac{K}{2}  \left(\nabla \phi({\bf r})\right)^2\right\rbrace
\nonumber\\
  \hat{\cal F}^{(\mathrm{GA})}[ c({\bf r})] &= 
k_B T\int d{\bf r}\, \left\lbrace 
\frac{r_0}{2}\phi({\bf r})^2 \right\rbrace
\label{GA}
\end{align}
The quadratic  models are analytically  solvable and they  serve as a
benchmark comparison for the results on the Ginzburg-Landau model.

The  bare  free  energy  function  $\hat{F}({\bf c})$  is  defined  in
(\ref{bareFdef}).     By   substituting    the    interpolated   field
$\overline{c}({\bf  r})=\psi_\mu({\bf r})c_\mu$ (repeated  indices are
summed over) one obtains
 \begin{align}
 \hat{F}^{({\rm GL})}({\boldsymbol{\phi}}) &=
k_BT
     \left\lbrace 
          \frac{r_0}{2}\phi_\mu {\bf M}_{\mu\nu}^\psi \phi_\nu 
        +\frac{K}{2}
        \phi_\mu {\bf L}_{\mu\nu}^\psi \phi_\nu
        \right\rbrace
\nonumber\\
&+ k_BT  \frac{u_0}{4}   F^{(4)}({\boldsymbol{\phi}}) 
\label{FGLdis0}
\end{align}
where  the   mass  matrix   ${\bf  M}^\psi_{\mu\nu}$  is   defined  in
(\ref{Mdelta})  and the  stiffness matrix  ${\bf L}^\psi_{\mu\nu}$  is
given by
\begin{align}
{\bf L}_{\mu\nu}^{\psi}&\equiv  \int d{\bf r}\, \nabla \psi_\mu({\bf r}) \cdot \nabla \psi_\nu({\bf r})
\label{ML}
\end{align}
and the quartic contribution is defined as
 \begin{align}
F^{(4)}(\phi)    &= M_{\mu\mu'\nu\nu'}^{\psi}\phi_{\mu}\phi_{\mu'}\phi_{\nu}\phi_{\nu'} 
\label{F4}
\end{align}
with a four-node mass given by
\begin{align}
M_{\mu\mu'\nu\nu'}^{\psi}&=  \int d{\bf r}\, \psi_\mu({\bf r}) \psi_{\mu'}({\bf r})\psi_{\nu}({\bf r}) \psi_{\nu'}({\bf r}) 
\label{M4}
\end{align}
Note   that   due   to    the   form   of   $\psi_\mu(\bf   r)$   (see
Fig.~(\ref{fig:delaunay1D}))  the  elements   of  the  matrices  ${\bf
  M}^\psi_{\mu\nu},{\bf L}^\psi_{\mu\nu}$ will be non-zero only if the
nodes $\mu,\nu$ coincide or are nearest neighbors.  In a similar way,
the elements  $ M_{\mu\mu'\nu\nu'}^{\psi}$ of the  four-node mass will
be  different  from  zero   only  if  $\left\lbrace  \mu,  \mu',  \nu,
\nu'\right\rbrace$ coincide or are all of them nearest neighbors.
\color{black}

The GL model shows phase  separation when $T<T_c$ giving concentration
fields that  have two distinct  values in different regions  of space.
In  the  present   paper,  though,  we  will   restrict  ourselves  to
supercritical  temperatures $T>T_c$  in such  a way  that there  is no
phase transition.   Note that  the statistics required  in subcritical
simulations  needs to  sample  the diffusion  of  the phase  separated
droplets, which  is usually very slow  \cite{Bray1994}.  In addition,
for supercritical temperatures translation  invariance leads to simple
forms for the  structure factor, which is the basic observable that
we will  consider in the present  paper.  

\section{Simulation results}
In this section, we consider a 1D periodic system governed by the free
energy functionals (\ref{eq:GLFEfunctional})-(\ref{GA}).   In 1D these
models are  well behaved, and  the continuum equations have  a precise
interpretation. We are concerned with the convergence of the numerical
method to the solution of the continuum equations as the grid is refined.
\subsection{Time discretization}
Up to  now we  have considered  the space discretization  of a  PDE or
SPDE,  where time  is  still  a continuum  variable.   Of course,  the
numerical resolution  of these  equations require a  discretization in
time. For the GL  model, there is a part in the SDE  that is linear in
the concentration  field and that  we call  the diffusive part  of the
SDE.  The  non-diffusive part arises from  the quartic term in  the GL
free energy. In order to be able  to use large time step sizes that do
not suffer from instabilities, we will treat the diffusive part of the
equation  implicitly, while  the  non-diffusive part  will be  treated
explicitly, following the implicit trapezoidal method proposed in Refs.
\cite{PhysRevE.87.033302,   MultiscaleIntegrators}.    This   temporal
integrator has the property that  for linear equations, when all terms
are   discretized  using   the  implicit   trapezoidal  rule,   it  is
unconditionally  stable   and  gives   the  same   static  covariances
(structure  factors) independent  of  the time  step size.   Therefore
temporal integration  errors in the  static factors are  eliminated by
this  scheme for  the GA  and GA+sigma  models.  When  some terms  are
discretized   explicitly,  as   for  the   GL  model,   some  temporal
discretization error will  be observed \cite{PhysRevE.87.033302}. Also
note that resolving  the correct {\em dynamic}  correlations for large
wavenumbers  requires   choosing  a   sufficiently  small   time  step
size. Note that the smallest  relaxation time is the one corresponding
to the wavenumber $k=\pi/a$ { where $a$ is the lattice spacing}. We use
a time  step smaller than this  relaxation time, but one  can use much
larger time steps  and still recover the correct  structure factor for
low wavenumbers since the algorithm is implicit.

\subsection{Parameters}
\label{sec:param}
The  set  of  parameters  in  the  van der  Waals  model  and  in  its
approximate form,  the Ginzburg-Landau  model, is the  following.  The
parameters corresponding to the particular fluid being studied are the
excluded volume  $b$ of  a van  der Waals  molecule, the  length scale
$\sigma$ of the potential, and  the critical
temperature  $T_c$ of  the  van der  Waals  fluid.  The  parameters
corresponding to the  thermodynamic state are the  temperature $ T$
and the global  concentration $c_0=N/L$ where $N$ is  the total number
of particles  and $L$ is  the size of  the box.  Because  the dynamics
conserves the total number of particles $N=\int d{\bf r}c_0({\bf r})$,
the total  number of particles is  a parameter of the  simulation that
enters through  the initial  conditions specified through  the initial
profile $c_0({\bf  r})$.  The  parameter corresponding to  the dynamic
equation is the mobility $\Gamma$ assumed  to be constant and given in
terms  of   the  diffusion  coefficient  $D$   as  $\Gamma=Dc_0/k_BT$.
Finally, we  have a set  of numerical  parameters, like the  time step
size $\Delta t$  and the total number  $M$ of nodes of  the mesh. Each
node has a volume ${\cal V}_\mu$ with $\sum^M_\mu {\cal V}_\mu = L$ .

From this  set of parameters, we  choose $b, k_BT_c, D$  as our units,
thus fixing the basic units of length, time, and mass. This results in
the  following dimensionless  numbers  as our  free parameters  $L/b$,
$\sigma/b$, $T/T_c$, $N$, $M$.  We will consider a fluid characterized
by a fixed  value of $b,\sigma,k_BT_c$.  In this way,  we will fix the
ratio $b/\sigma=10$.  We  also fix $N$ in order to  have the total
concentration $N/L$  equal to the critical  concentration $1/3b$, this
is, $N=L/3b$. In this way, the number of free parameters to explore is
reduced to $L/b$, $M$, $T/T_c$.   The limit $L/b\to\infty$ is the {\em
  thermodynamic  limit} or  infinite  system size  limit, whereas  the
limit $Mb/L\rightarrow \infty$  (so the volume of  each cell approaches
zero) is  the {\em continuum  limit}. 

In the following sections, all the  simulations are performed at a box
of size $L=10$ at a temperature $k_B  T = 1.11$ in the selected units,
with  the corresponding  parameters  in Eq.  (\ref{param}) being  $r_0
\simeq 0.07$  and $K \simeq  0.007$.  They  all start from  an initial
state  in which  $c_{\mu}(t=0)  = c_0$  for all  $\mu$,  and employ  a
sufficiently  small  time  step  to  ensure  numerical  stability  and
convergence  of  results.   We   ensure  that  we  sample  equilibrium
configurations by compiling statistics only  after a time of the order
$L^2 / D$.  The number of  particles $N = \sum_\mu c_\mu {\cal V}_\mu$
is exactly conserved by the algorithm.
\subsection{Observables}
The structure  factor is an  observable that is specially  suited when
there is translation invariance.  The  structure factor is the discrete
Fourier transform of the matrix of covariances, this is, the matrix of
second moments  of the probability distribution  $P^{\rm eq}({\bf c})$
in  Eq.   (\ref{Einsdisc})  (see Appendix  \ref{app:structure}).   The
$k$-dependent structure  factor allows to discuss  correlations of the
concentration at different length scales.  The structure factor can be
analytically computed in  the continuum limit for  a GA+$\sigma$ model
as shown in Appendix \ref{app:structure}, with the result
\begin{align}
S^c(k) &
= \left\langle \delta c({ k},0) \delta c(-{ k},0)\right\rangle 
= \frac{c_0^2}{r_0} \frac{1}{1+ \frac{k^2}{k_0^2}} 
\label{sck}
\end{align}
where 
\begin{align}
k_0 = \left(\frac{r_0}{K}\right)^{1/2}=\frac{1}{\sigma}  \left(\frac{T}{T_c}-1\right)^{1/2}
\end{align}
The typical length scale below which fluctuations start to decorrelate is
given by $\lambda=2\pi/k_0$.

The \textit{dynamic} structure factor is  the Fourier transform of the
time  dependent  correlation  function  and  can  also  be  explicitly
computed for the Gaussian model leading to
\begin{align}
S^c(k,t) &= \left\langle 
\delta \hat{c} ({ k}, t) \delta \hat{c} (-{ k},0)
\right\rangle
=
S^c({ k}) \exp 
\left\lbrace
- \frac{t}{\tau_k}
\right\rbrace
  \label{eq:SKtGAs}
\end{align}
with a typical relaxation time given by
\begin{align}
  \tau_k &= \left[ \frac{D}{c_0} r_0 \left(1+ \frac{k^2}{k_0^2} \right) k^2\right]^{-1}
\label{tauk}
\end{align}
The continuum results (\ref{sck}) and  (\ref{tauk}) serves also as the
basis for  computing the structure  factors of the  discrete variables,
see  Appendix  \ref{app:structure}.   

In  addition  to  the  structure  factor, we  will  also  consider  as
observable the  probability that  a region  of finite  size $l$  has a
given number  of particles  in its interior.   In 1D,  this observable
should be  independent of the  resolution, given a  sufficiently large
resolution, and will  allow us to detect whether the  GL model behaves
in a Gaussian or non-Gaussian way, depending on the temperature.

\subsection{Regular lattice results}

\subsubsection{Static structure factor for Gaussian models}

While  the structure  factor (\ref{sck})  has an  explicit expression,
what we compute in a  simulation is the covariance $\langle \delta
c_\mu\delta c_\nu\rangle$  of the  discrete variables $c_\mu$ or, for
regular lattices, its Fourier transform. We
introduce the  discrete Fourier transform $  \hat{c}_m$ with $m=0,M-1$
of the discrete concentration field $c_\mu$ according to
\begin{align}
  \hat{c}_m&=\frac{1}{M}\sum_{\mu=0}^{M-1} e^{-i\frac{2\pi}{L} mr_\mu}c_\mu
\label{DFFT0}
\end{align}
and define the discrete structure factor as \cite{Donev2010}
\begin{align}
  \hat{S}^c(k)&\equiv L\left\langle \delta \hat{c}_m \delta \hat{c}^*_{m} \right\rangle
\end{align}
where $k=\frac{2\pi}{L}m$ for integer $m$.  
The  modes $\hat{c}_m$  are related  to  $c_\mu$ which,  in turn,  are
related to the  continuum field through Eq.   (\ref{natural}). For the
GA+$\sigma$ model we know the  correlations of the fluctuations of the
continuum field and, therefore, we have an explicit expression for the
discrete structure factor (see Appendix \ref{app:structure})
\begin{align}
\hat{S}^c(k)&=\frac{c_0^2}{r_0} \frac{9}{\left[2+\cos\left(ka\right)\right]^2}
\sum_{\alpha\in\mathbb{Z}}
\frac{\mathrm{sinc}^4\left(\frac{ka}{2}-\pi\alpha\right)}
{1+\left(\frac{k}{k_0}-\frac{2\pi \alpha }{k_0a}\right)^2}
\label{ScFIN}
\end{align}
 Note that in the limit of
high resolution  $a=L/M\to 0$, the  only term that contributes  in the
sum over  $\alpha$ is $\alpha=0$.   In this limit, then,  the discrete
structure factor  (\ref{ScFIN}) converges towards the  continuum limit
(\ref{sck}).  Eq.  (\ref{ScFIN}) gives the prediction of the continuum
theory   for  the   covariance   of  fluctuations   of  the   discrete
concentration variables.

\begin{figure}[!tb]
    \includegraphics{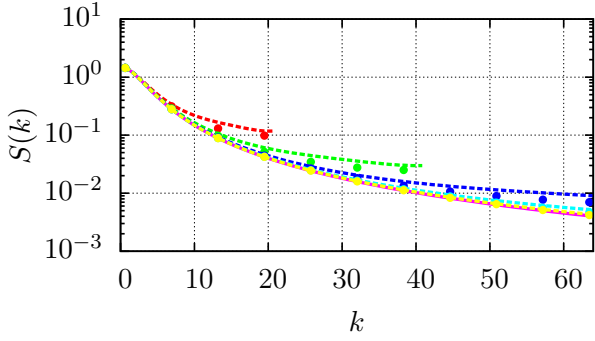}
    \caption{\label{fig:GL_SK_comparison}    Comparison     for    the
    GA$+\sigma$  model  of  the   static  structure  factors  $S^c(k)$
    (dashed  lines) in  Eq.  (\ref{ScFIN}),  $S^d(k)$ (points)  in Eq.
    (\ref{SdFIN}),  and $S(k)$ (solid  pink line) in Eq.  (\ref{sck}).
    From top to bottom:  red, $M=64$;  green, $M=128$,  blue, $M=256$;
    cyan,  $M=512$;   yellow,  $M=1024$;   pink,  continuum  structure
    factor.  As the resolution increases,  the  range of $k$ for which
    there is  no significant discrepancy between  the discrete results
    and the continuum prediction $S(k)$ in Eq.  (\ref{sck}) increases.
    }
\end{figure}

The          numerical         integrator          proposed         in
\cite{PhysRevE.87.033302,Donev2014} produces the same static structure
factor  regardless  of  the  time   step.   As  we  show  in  Appendix
\ref{app:DiscreteSKNum},   the   actual   discrete   structure  factor
$\hat{S}^d(k)$ produced by our integrator for the GA+$\sigma$ model is
given by
\begin{align}
\hat{S}^d(k)
&=\frac{c_0^2 }{r_0}\frac{3}{\left[2+\cos\left(ka\right)\right]}\frac{1}{1+\frac{k^2}{k_0^2}\left(\frac{3 \mathrm{sinc}^2(ka/2)}{(2+\cos ka)}\right)}
\label{SdFIN}
\end{align}
which is  independent of  the time  step $\Delta  t$ \cite{Donev2014}.
This result (\ref{SdFIN}) is useful as  it allows to check for correct
coding of  the algorithm. We  have indeed verified that  the numerical
results lead  exactly to  (\ref{SdFIN}).  Note that  $\hat{S}^d(k)$ in
(\ref{SdFIN}) tends to the continuum limit $\hat{S}(k)$ in (\ref{sck})
for     $k<<\frac{\pi}{a}$.     In     the    limit     $k_0\to\infty$,
$\hat{S}^d(k)=\hat{S}^c(k)$ (see Eq.  (\ref{SckGA})). For finite $k_0$,
$\hat{S}^d(k)$  is   different  from  $\hat{S}^c(k)$,   although  both
structure  factors  tend  to  the  continuum  value  $\hat{S}(k)$  for
sufficiently    high     resolutions.     We    compare     in    Fig.
\ref{fig:GL_SK_comparison}    $\hat{S}(k)$   in    Eq.    (\ref{sck}),
$\hat{S}^c(k)$  in  Eq.   (\ref{ScFIN})   and  $\hat{S}^d(k)$  in  Eq.
(\ref{SdFIN})  for   increasing  levels   of  resolution.    The  main
observation  is  that  $\hat{S}^d(k)$   and  $\hat{S}^c(k)$  are  very
similar.   In other  words,  not only  the  infinite limit  resolution
$\hat{S}(k)$ is  well captured by  the numerical method, but  also the
predictions of the continuum theory for a finite mesh are equally well
reproduced.

\begin{figure}[t]
    \includegraphics{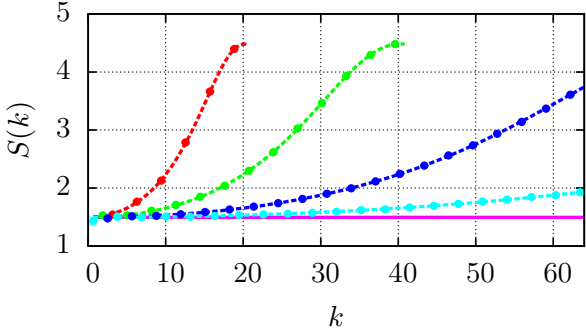}    
    \caption{\label{fig:GA_Sk}Static structure factor as a function of
    $k$  for the  model GA.  From  left to  right:  red,  $M=64$ nodes
    ($\Delta t = 10^{-3}$);  green,  $M=128$  ($\Delta t = \frac{1}{4}
    10^{-3}$);  blue,  $M=256$ ($\Delta  t  =  \frac{1}{16} 10^{-3}$);
    cyan,  $M=512$  ($\Delta t  = \frac{1}{32}  10^{-3}$);  pink solid
    line,  continuum  result $c_0^2/r_0$  given by  Eq.~(\ref{sck}) in
    the  limit  $k_0\to\infty$.   Dots  correspond  to  the  numerical
    structure  factor   obtained   from   simulations;   dashed  lines
    correspond    to    the    theoretical    prediction    given   by
    Eq.~(\ref{ScFIN}).}
\end{figure}
The Gaussian model GA is obtained by setting $K=0$ and suppressing the
square gradient term. This implies $k_0=\infty$ in Eq.  (\ref{GA}) and
results in that different points in space are completely uncorrelated.
Figure~(\ref{fig:GA_Sk})  shows the  \textit{static} structure  factor
for  different resolutions,  from $M=64$  to  $M=256$ as  well as  the
continuum    solution    \cite{Donev2010}.     We    also    plot    in
Fig.~(\ref{fig:GA_Sk})  the  theoretical  discrete  structure  factor,
given by Eq.~(\ref{ScFIN}),  which takes into account  the finite size
of the  cell. The  simulation results  are indistinguishable  from the
theoretical prediction at  each resolution as they must  since in this
case  Eq.  (\ref{ScFIN})  is  equal  to  (\ref{SdFIN}).   As  we  keep
increasing  the resolution,  the  range of  wavenumber  for which  the
structure  factor coincides  with  the prediction  $c_0^2/r_0$ of  the
continuum theory increases.  However, there is always a discrepancy at
large wavenumbers corresponding to the inverse of the lattice spacing.

It  should be  mentioned that  the analytic  results obtained  for the
correlation  of the  discrete concentration  of nodes  $\langle \delta
c_\mu\delta  c_\nu\rangle$  in  the Appendix  \ref{app:structure}  are
based on  the canonical ensemble.  Therefore, they do not  satisfy the
sum    rule    $\sum_\mu{\cal   V}_\mu\langle    \delta    c_\mu\delta
c_\nu\rangle=0$ that results from the conservation of the total number
of  particles.   The latter  property  is  actually satisfied  by  the
simulation results.   The differences, however, are  vanishingly small
in the thermodynamic limit.

\subsubsection{Dynamic structure factor for Gaussian models}

The  \textit{dynamic}  structure  factor  can also  be  obtained  from
Eq.~(\ref{eq:SKtGAs}) for a given $k$ value. 
\begin{figure}[t]
    \includegraphics{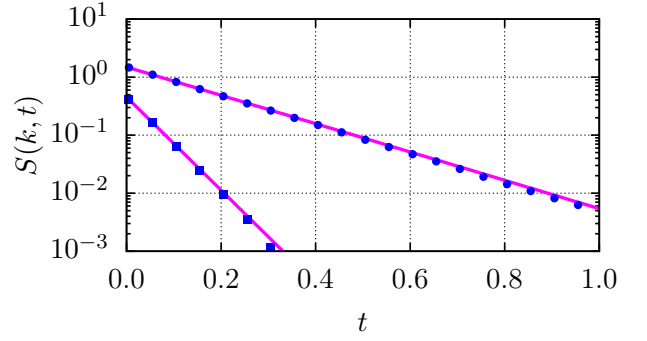}    
    \caption{\label{fig:GA_k_t} Dynamic structure  factor for $k=5.02$
    as a  function of time for  the model GA  (blue circles,  top) and
    GA$+\sigma$   (blue   squares,    bottom).    Averaged   over   10
    simulations at  $M=256$.  Circles  and  squares  correspond to the
    numerical result,  solid pink line  corresponds to the theoretical
    prediction  (\ref{eq:SKtGAs}).   In  the  GA  model,  $\sigma^2=0$
    gives  a  relaxation  time  $\tau_k  =  0.2$.  In  the GA$+\sigma$
    model,  $\sigma^2=0.01$ ($K\simeq 0.007$)  gives a relaxation time
    $\tau_k=0.05$.}
\end{figure}
Figure~(\ref{fig:GA_k_t})  shows  the  dynamic  structure  factor  for
$k=5.02$  with $M=256$  (a sufficiently  fine  grid) for  both the  GA
(circles) and the GA$+\sigma$ (squares) models, and compares numerical
results with the theoretical prediction  (pink solid line).  In the GA
case, the value $r_0 \simeq 0.07$ gives a relaxation time of $\tau_k =
0.2$.  In the GA$+\sigma$ case,  the parameter $r_0$ remains unchanged
and $K\simeq 0.007$, with a time scale $\tau_k \simeq 0.05$. As can be
seen, numerical simulations overlap  with theoretical predictions.  We
also plot in Fig.~\ref{fig:GAs_SK_t_tau}  the relaxation time $\tau_k$
obtained through simulations for both the GA (circles) and GA$+\sigma$
(squares)   models,   and   compares   them   with   the   theoretical
result~(\ref{tauk}).  Both  results overlap  the theoretical  ones for
time  scales  smaller  than  $10^{-4}$  in  reduced  units,  which  is
comparable to  the time  step size $\Delta  t =  \frac{1}{16}10^{-3} =
6.25 \times10^{-5}$. Note that this time step is much smaller than the
relaxation time  for the  wavenumber plotted. We  may still  have good
results for small wavenumbers with much larger time steps, but we have
decided  to use  a  time step  that would  resolve  also the  smallest
relaxation   times,   which    is   roughly   $\tau_{\rm   min}=\Delta
x^2/D=L^2/(M^2 D)=1.5\times 10^{-3}$.

\begin{figure}[t]
    \includegraphics{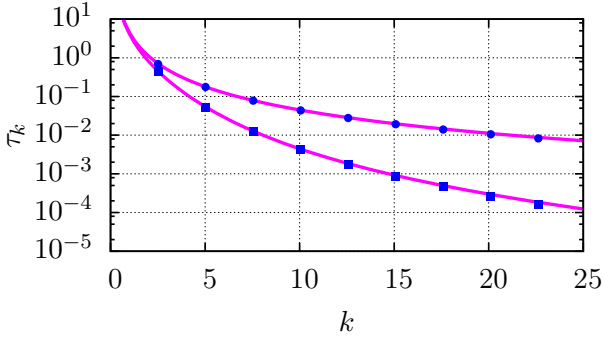}
    \caption{\label{fig:GAs_SK_t_tau} Relaxation  time  $\tau_k$  as a
    function of  $k$ for $M=256$ in  both the GA  model (blue circles,
    top) and  the GA$+\sigma$ model  (blue squares,  bottom) averaging
    over  10  simulations with  time  step  $\Delta  t  = \frac{1}{16}
    10^{-3}$.  Dots correspond  to the  relaxation time  obtained from
    a  numerical  fitting  of  the  dynamic  structure  factor  to  an
    exponential  function.   Lines   correspond  to   the  theoretical
    prediction in Eq.  (\ref{tauk}).}
\end{figure}

\subsubsection{Static structure factor for Ginzburg Landau model}
Once the code has been checked for the Gaussian models, we may move to
the   more    interesting   case   of    the   Ginzburg-Landau   model
Eq. (\ref{eq:GLFEfunctional})  with its discrete  free energy function
given   in  Eq.~(\ref{eq:GLFEdiscrete}).    This  model   shows  phase
separation  at   subcritical  temperatures.   For   sufficiently  high
supercritical temperatures Gaussian  behavior is recovered.  In order
to detect  interesting non-linear effects, albeit in  the single phase
region,  we  will  explore  temperatures  near  (above)  the  critical
temperature characterized by a single non Gaussian phase.

\begin{figure}[t]
    \includegraphics{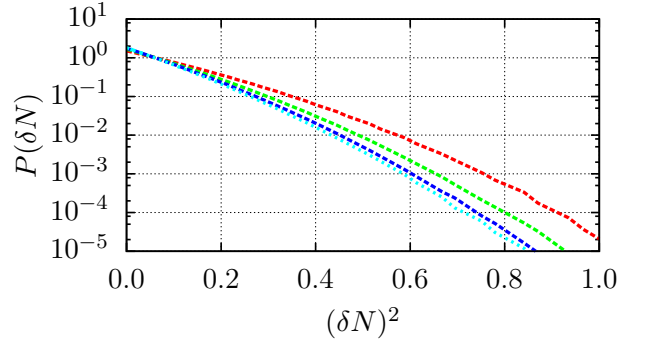}
    \caption{\label{fig:GL_HIST_overKTC}  Probability  distribution of
    finding a deviation  from  the  mean  of  the  number of particles
    inside a  region of size  $l = \frac{1}{16}  L$ for the  GL model.
    This  is,  $\delta  N  =  c_0  l  -  \sum_{\mu\in  l}  c_\mu {\cal
    V}_\mu$.  From top to bottom,  $M=64$ nodes,  $M=128$, $M=256$ and
    $M=512$.  We observe  convergence of the  probability distribution
    towards  a   non  Gaussian  distribution  as   the  resolution  is
    increased.  }
\end{figure}

Fig.~(\ref{fig:GL_HIST_overKTC}) shows the probability distribution of
finding a deviation from the mean  of the number of particles, $\delta
N$, inside a region of size  $l = \frac{1}{16}L$.  The simulation were
done at  $k_BT = 1.11$ ($r_0  \simeq 0.07$) and $\sigma^2  = 0.01$ ($K
\simeq 0.007$) in  the selected units.  As we  increase the resolution
the probability distribution  converges towards a unique  limit.  In a
Gaussian model, one should expect a linear dependence between $(\delta
N)^2$ and  $P(\delta N)$. This is  not observed in the  limit curve of
Fig.~(\ref{fig:GL_HIST_overKTC}), signaling non-Gaussian behavior for
this thermodynamic point state.

\begin{figure}[t]
    \includegraphics{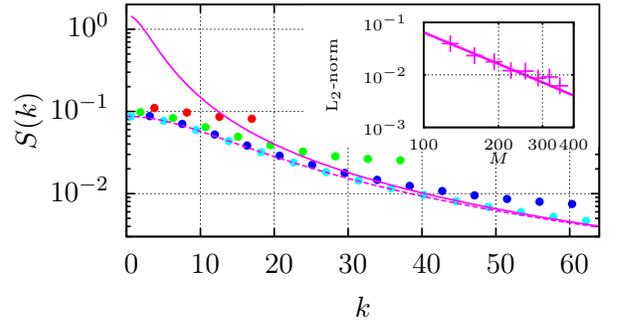}    
    \caption{\label{fig:GL_Sk}Static structure factor as a function of
    $k$ for the GL model.  Red,  $M=64$  nodes ($\Delta t = 10^{-3}$);
    green,  $M=128$ ($\Delta t = \frac{1}{4} 10^{-3}$);  blue, $M=256$
    ($\Delta t = \frac{1}{16}  10^{-3}$);  cyan,  $M=512$ ($\Delta t =
    \frac{1}{32} 10^{-3}$).  Convergence  of the numerical  results is
    observed as the resolution  increases.  With solid pink line,  the
    continuum  structure  factor of  the  GA$+\sigma$  model  with the
    same parameters $r_0  \simeq  0.07$  $K  \simeq  0.007$  as the GL
    model.  Dashed pink  line shows the continuum  structure factor of
    a renormalized  GA$+\sigma$ model which  has the same  variance as
    the GL model.  The  empirical  fitting  of  the  numerical data to
    the  renormalized   GA$+\sigma$  static  structure   factor  gives
    $r_0=1.27$   and   $K=0.007$.    Inset,    L$_2$-norm   indicating
    convergence.  }
\end{figure}

Figure~(\ref{fig:GL_Sk})  shows the  static structure  factor for  the
Ginzburg-Landau model at different  resolutions, $M=64$ (red), $M=128$
(green) and $M=256$ (blue) and $M=512$  (cyan).  We observe that as we
increase  the resolution  we converge  towards a  unique answer.   The
L$_2$-norm      L$_2$$(M_1,M_2)=\sqrt{\sum_i      (S^{M_1}(k_i)      -
  S^{M_2}(k_i))^2}$    is    also    shown    in    the    inset    of
Fig.(\ref{fig:GL_Sk}), where we compare  the structure factor obtained
at resolution $M_1$ with the one obtained at a higher resolution
$M_2=M_1+32$.  A  pink line of  slope -2 agrees well with the numerical results reflecting
second order spatial convergence of the algorithm.

We also compare in  Fig.~(\ref{fig:GL_Sk}) the static structure factor
of the GL  model with the continuum limit of  the corresponding one in
the GA$+\sigma$ model. Two regions  are clearly observed, separated by
a value at around $k_c=30$.
On one hand,  for $k  < k_c$
(large length scales) there is a clear difference between the Gaussian and
the GL  model.  For  small wavenumbers, the  contribution of  the quartic
term  is important  and suppresses  the amplitude  of the  fluctuations
relative to the  GA+$\sigma$ model.  On the other hand,  for $k > k_c$
there is  no difference between both  models in the limit  of infinite
resolution, and the quartic term  has a minimal effect.  The existence
of two  regions may be  understood from  the probability of  finding a
particular Fourier mode $\phi_k$ of the  field, which will be given by
the   exponential  of   the  free   energy  (\ref{eq:GLFEfunctional}),
expressed in  Fourier space.  The  quadratic term in this  free energy
has  a $k$-dependent  prefactor  $(r_0+K k^2)/2$.   Near the  critical
point, we have $r_0\sim0$.  Therefore,  for $k\sim 0$, the free energy
is entirely  dominated by  the quartic  interaction (which  in Fourier
space is  in the form of  a convolution).  At sufficiently  large $k$,
however, the quadratic term dominates over the quartic.  The effect of
the  quartic  term  is  to  strongly suppress  the  amplitude  of  the
long-wave fluctuations  with respect  to the  Gaussian model  with the
same  $r_0,K$ parameters.

\subsubsection{Dynamic structure factor for Ginzburg Landau model}

\begin{figure}[t]
    \includegraphics{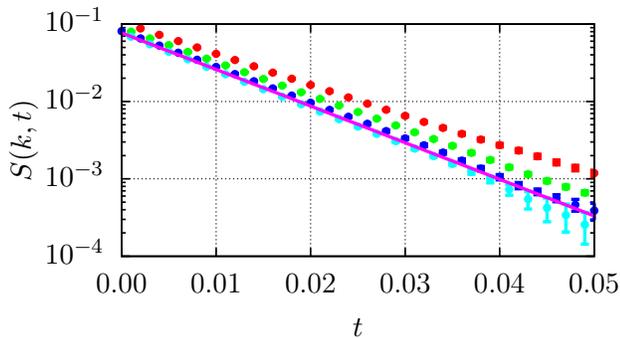}
    \caption{\label{fig:GL_Sk_k314_t}Dynamic  structure  factor  as  a
    function of $t$  for $k=5.02$ for the  GL model.  With dots:  red,
    $M=64$ nodes ($\Delta t = 10^{-3}$);  green,  $M=128$ ($\Delta t =
    \frac{1}{4} 10^{-3}$);  blue,  $M=256$  ($\Delta t  = \frac{1}{16}
    10^{-3}$).  With solid pink line,  the dynamic structure factor of
    a renormalized  GA$+\sigma$ model  with parameters  $r_0=1.27$ and
    $K=0.007$.  }
\end{figure}

\begin{figure}[t]
    \includegraphics{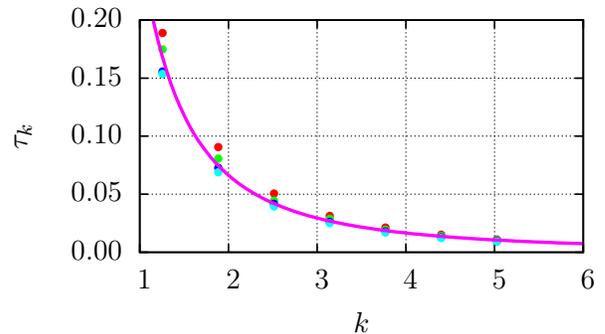}
    \caption{\label{fig:GL_SK_tau_all}Relaxation  time  $\tau_k$  as a
    function of  $k$ in the  GL model at  different resolutions $M=64$
    (red),  $M=128$ (green),  $M=256$ (blue)  and $M=512$ (pink).  All
    of them  obtained averaging over  10 simulations.  Dots correspond
    to  the  numerical  relaxation  time  (obtained  from  a numerical
    fitting  of  the  dynamic   structure  factor  to  an  exponential
    function).   Line  corresponds   to  the   theoretical  prediction
    (\ref{tauk}) of  the renormalized GA$+\sigma$  model with $T=-1.4$
    ($r_0 \simeq 1.27$) and $\sigma^2 = 0.01$ ($K\simeq 0.007$).}
\end{figure}

Figure~(\ref{fig:GL_Sk_k314_t}) shows the  dynamic structure factor of
the  GL  model for  $k=5.02$  at  different resolutions.   We  observe
convergence as  the resolution  is increased in  the region  where the
statistical errors  are small ($S(k,t)  \sim 10^{-3}$). The  fact that
the  decay  of  the  dynamic  structure factor  of  the  GL  model  is
exponential suggests  that its dynamics is  very similar to that  of a
\textit{renormalized}  Gaussian   model.   In   order  to   test  this
conjecture, we have  considered the best GA+$\sigma$  model that would
reproduce the \textit{static}  structure factor of the  GL model.  The
best Gaussian model  is the one that has the  same structure factor as
that of  the GL  model.  The result  of the fit  is presented  in Fig.
\ref{fig:GL_Sk} and gives  the parameters $r_0 =  1.27$ and $K=0.007$.
Observe that in the renormalized GA+$\sigma$ model the surface tension
coefficient  $K$ is  the  same and  only the  value  of the  quadratic
coefficient  $r_0$ is  renormalized,  consistent  with predictions  of
renormalization  (perturbative)  theories  \cite{HairerReview}.   With
these values of $ r_0 ,K$  we compute independently the prediction for
the  relaxation time  given  by Eq.   (\ref{tauk})  for a  GA+$\sigma$
model.  The result is the solid line in  \ref{fig:GL_SK_tau_all}.
A very good agreement between the  measured relaxation times of the GL
model  and  the prediction  of  this  renormalized Gaussian  model  is
obtained.   This suggests  that  indeed,  the GL  model  behaves as  a
GA+$\sigma$ model with renormalized  parameters.  

\subsection{Irregular lattices}

In this section, we present similar results as in the previous section
but in  this case  for irregular  lattices.  Adaptive  mesh resolution
allow one  to resolve interfaces appearing  below critical conditions,
and deal with complicated boundary  conditions.  In the present paper,
while we  still remain in  the supercritical  region of the  GL model,
where  no  interfaces are  formed,  we  test  the performance  of  the
algorithm  presented for  irregular lattices.   We consider  irregular
lattices constructed  by displacing  randomly the  nodes of  a regular
lattice,  allowing for  a maximum  fluctuation  of   $\pm$ 40\%  with
respect to  the regular lattice configuration.   These random lattices
are  a worst  case scenario  and  other lattices  with slowly  varying
density of nodes behave much better in terms of numerical convergence.
We compare regular  and irregular lattice simulation  results by using
the same set of parameters in  both cases.  Typically, what we observe
is that higher resolutions are required in irregular lattices in order
to achieve comparable accuracy as those in regular lattices.  The time
step  in an  irregular lattice  is  dictated by  the shortest  lattice
distance $ \Delta x_{\rm min}$ encountered according to $\Delta t \sim
\Delta x^2_{\rm min}/D$.

From a numerical point of  view, obtaining the static structure factor
for regular grids can be efficiently done with Fast Fourier Transforms
(FFT): we  just need to perform  a FFT of the  concentration field and
multiply  it  by  its  complex conjugate.   However,  irregular  grids
complicate the use of the FFT and  we need to follow a different route
to obtain the static structure factor.  The idea is to interpolate the
discrete field on  the irregular coarse grid onto a  very fine regular
grid  on which  the  FFT can  be used.  Of  course, the  interpolation
procedure  modifies  the  structure  factor because  we  are  creating
information at the interpolated points.

At the  same time, when  we consider irregular  grids, we do  not have
simple analytical results to compare, even for the Gaussian models. In
this  case,  our strategy  is  to  produce synthetic  Gaussian  fields
generated in  a very fine grid  ensuring that they are  distributed in
such a way that have a structure factor given by (\ref{sck}).  This is
achieved by generating  random Gaussian numbers in  Fourier space with
the correct mean  and covariance for each wavenumber $k$ so that the
theoretical $S(k)$  is recovered. These synthetic  Gaussian fields are
taken as the  ``truth'' to compare with.  From  the synthetic Gaussian
field, we compute  a coarse-grained field on an  irregular coarse grid
by applying  the coarsening operator  $\delta_\mu({\bf r})$ as  in the
first equation (\ref{natural}), where  the integral is approximated as
a  sum over  the very  fine  grid.  This  gives us  realizations of  a
Gaussian  field in  a  coarse irregular  grid. We  may  now apply  the
methodology used for computing the  structure factor in regular grids,
by interpolating on a very fine regular grid and using the FFT.

\begin{figure}[h!t]
    \includegraphics{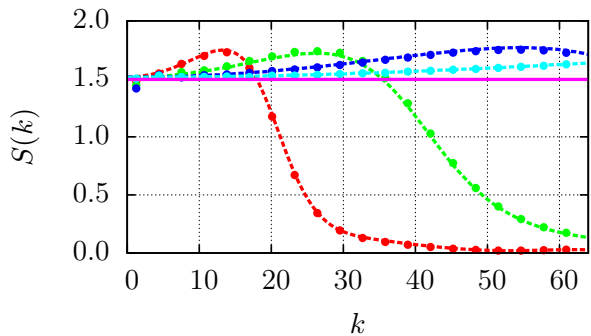}
    \caption{\label{fig:GA_Sk_irr}Static   structure   factor   as   a
    function of $k$  for  the  model  GA  in irregular lattices.  From
    left  to right,  $M=64$  nodes  ($\Delta  t  = 10^{-3}$),  $M=128$
    ($\Delta  t  =  \frac{1}{4}   10^{-3}$),   $M=256$  ($\Delta  t  =
    \frac{1}{16}  10^{-3}$),   $M=512$   ($\Delta  t   =  \frac{1}{32}
    10^{-3}$) and  continuum limit of  the GA model  (solid pink line,
    Eq.~\ref{sck}).   Dots  correspond  to   the  simulations  of  the
    diffusion  equation,   while   dashed  lines  correspond   to  the
    synthetic  Gaussian  fields.  The  striking  difference  with Fig.
    \ref{fig:GA_Sk} is  due  to  the  interpolation  procedure used to
    compute the static structure factor in the irregular grid.}
\end{figure}

\begin{figure}[h!t]
    \includegraphics{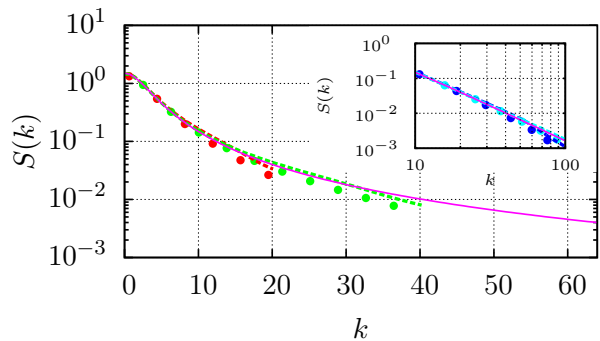}
    \caption{\label{fig:GAs_Sk_irr}    Structure    factor    in   the
    GA+$\sigma$ model  for an irregular  lattice.  In the  main panel,
    $M=64$ nodes  (red,  $\Delta  t  =  10^{-3}$)  and $M=128$ (green,
    $\Delta  t  = \frac{1}{4}  10^{-3}$).  In  the  inset  we  plot in
    log-log  scale  results  for  $M=256$  nodes  (blue,  $\Delta  t =
    \frac{1}{16}  10^{-3}$)  and  $M=512$  nodes  (cyan,  $\Delta  t =
    \frac{1}{32}  10^{-3}$).  Dots correspond  to  the  simulations of
    the   diffusion  equation.   Dashed   lines   correspond   to  the
    synthetic  Gaussian  (with   surface  tension  term)  field.   The
    theoretical  prediction  in  Eq.~(\ref{sck})  is  also  plotted in
    solid pink line.  }
\end{figure}

Figures~(\ref{fig:GA_Sk_irr})  and  (\ref{fig:GAs_Sk_irr})  show,  for
both a GA  and a GA$+\sigma$ model, the  agreement between simulations
(in dots) and the synthetic procedure (dashed lines). We also show the
predictions obtained from (\ref{sck}), demonstrating that we correctly
discretized Eq. (\ref{PDE}) on the irregular grid.

\begin{figure}[t]
    \includegraphics{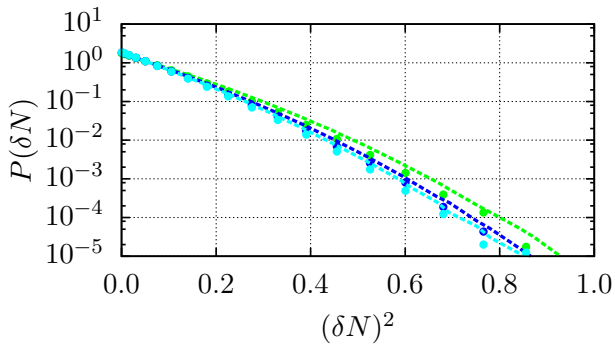}
    \caption{\label{prob_irr1}Probability  of $\delta  N$  in  a given
    region of space  $l= \frac{1}{16}L$ in a  random grid.  The points
    that limit the  region  are  kept  fixed  as  in the regular grid,
    inside  the region  the  nodes  are  randomly distributed.  Green,
    $M=128$;   blue,   $M=256$;   cyan,   $M=512$.   We   compare  the
    probability  in  a  random  grid  (points)  with  the  probability
    corresponding to a  regular grid with the  same resolution (dashed
    lines).}
\end{figure}

We move now to the GL  model. We consider the probability distribution
of a fluctuation of the number of particles in a fixed region of space
for the GL model.  The region of space is delimited  by two nodes that
are always at  the same distance $l=L/16$.  In a  first simulation, we
consider an arbitrary grid of nodes set at random in the whole domain,
except for the  two points delimiting the region of  interest that are
always fixed. In Fig. \ref{prob_irr1} we plot the result of increasing
the number of nodes in the simulation.

\begin{figure}[h!t]
    \includegraphics{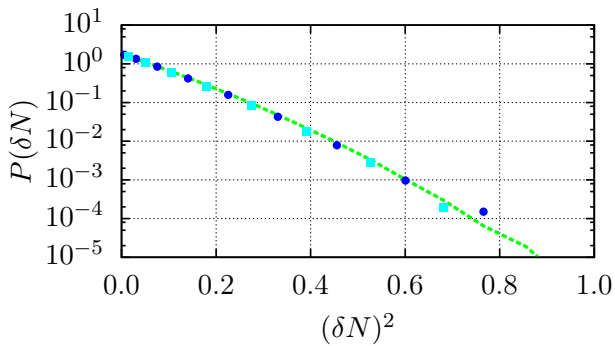}
    \caption{\label{prob_irr2}Probability  of $\delta  N$  in  a given
    region   of   space   $l=   \frac{1}{16}L$.   Green   dashed  line
    corresponds  to  a  regular   lattice  with  $M=128$.   Blue  dots
    correspond to  a grid with $M=256$  and cyan dots  correspond to a
    grid with $M=512$.  For the  $M=512$ grid,  64 nodes are uniformly
    distributed in half  of the box while the  remaining 448 nodes are
    distributed uniformly in the other  half.  In this way,  we have a
    grid  which  is,  in  one  region,  seven  times  finer  than  the
    original one;  in  the  other  region,  exactly  the original one.
    The grid  $M=256$ is defined  with $32$ nodes  in half of  the box
    and $224$ nodes in the other half.}
\end{figure}

In a second simulation, we divide  the box in 16 equally spaced regions
delimited by nodes of  the grid. Then, in half of the  boxes we have a
coarse resolution  and in the other half  we have a finer
resolution.  The  probability in  any of the  regions is
essentially  the  same,  as  shown in  Fig.  \ref{prob_irr2},  further
validating the method for irregular grids.

\begin{figure}[!htb]
    \includegraphics{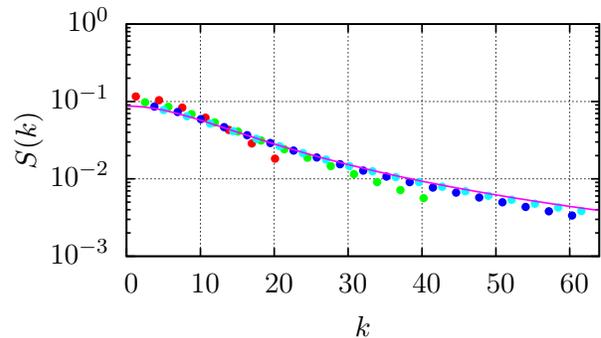}
    \caption{\label{fig:GL_irr_SK}  Static   structure  factor   as  a
    function of  $k$ for the  GL model in  an irregular lattice.  From
    bottom to top:  red,  $M=64$ ($\Delta  t = \frac{1}{16} 10^{-3}$);
    green,   $M=128$  ($\Delta  t  =  \frac{1}{32}  10^{-3}$);   blue,
    $M=256$  ($\Delta  t =  \frac{1}{64}  10^{-3}$)  and  cyan $M=512$
    ($\Delta t  =  \frac{1}{128}  10^{-3}$).  From  $M=64$ to $M=256$,
    averaged  over   10  simulations.   Pink  solid   line  shows  the
    theoretical  renormalized GA$+\sigma$  model (with  $r_0=1.27$ and
    $K=0.007$).}
\end{figure}

Finally,  we show  in  Fig. \ref{fig:GL_irr_SK}  the static  structure
factor  for the  GL  model  in an  irregular  random  grid, where  the
simulations  are  performed  with  the same  parameters  as  those  in
Figs.~(\ref{fig:GA_Sk_irr})  and  (\ref{fig:GAs_Sk_irr}).  We  observe
that  by  increasing the  resolution  the  structure factor  converges
towards a continuum  result, consistent with the results  based on the
regular  grid.   We conclude  that  the  algorithm presented  displays
convergence of the GL model for both regular and irregular grids.

\section{Conclusion}
We  have   proposed  a  Petrov-Galerkin  Finite   Element  Method  for
discretizing non-linear diffusion SPDE on arbitrary grids in arbitrary
dimensions. The method uses the concept of mutually orthogonal sets of
discretization  basis  functions  $\boldsymbol{\delta}({\bf  r})$  and
continuation basis functions $\boldsymbol{\psi}({\bf r})$.  We use two
different  basis functions  for what  is known  in the  finite element
literature  as trial (or  test) functions  and  solution functions.  As
opposed  to  a  Galerkin  method  in   which  the  weak  form  of  the
differential equation is constructed with $\boldsymbol{\psi}({\bf r})$
itself, the Petrov-Galerkin  method leads to a  positive semi definite
diffusion matrix.   This property is  crucial for representing  at the
discrete  level the  Second Law  satisfied by  the original  PDE. More
importantly, the  diffusion matrix  needs to  be positive  definite if
thermal fluctuations are to be introduced in the equation, because the
covariance  matrix  of   the  random  terms  is   just  the  diffusion
matrix. The method is general,  valid for regular and irregular grids,
and in any dimension.

Our approach combines mathematical  aspects of the discretization of a
PDE as  well as attention  to the physical  origin and meaning  of the
PDE. In fact, a given PDE  may be written in many different equivalent
forms by just reshuffling derivatives and functions.  For example, the
GA+$\sigma$ model for  constant mobility leads to a  PDE that contains
fourth order space  derivatives that may be written  in very different
ways.  We use
in  a rational  way  the physical  information  of the  origin of  the
different terms in order to  propose the discretization of the PDE. 

We have discussed the microscopic  foundation of the different dynamic
equations considered  in this  paper in order  to have  a well defined
physical  interpretation  for them.   The  selection  of the  relevant
variables (\ref{hatcmu})  used in the  ToCG to derive the  discrete SDE
has been guided by the finite element methodology, with a view to have
well defined  continuum limits when  possible.

We  have shown  that the  free  energy function  entering the  dynamic
equation for \textit{the ensemble average discrete concentration field} derived
from the ToCG,  and that entering the finite  element discretization of
Eq.  (\ref{PDE}), are actually the same  function, in the limit of high
resolution.   While  a  continuum  equation  like  (\ref{PDE})  has  a
well-defined meaning, its stochastic counterpart (\ref{SPDE}) obtained
from the underlying microscopic dynamics  has physical meaning only in
a discrete setting where, instead of an equation like (\ref{SPDE}) one
needs  to consider  an equation  for  discrete variables  of the  form
(\ref{eq:SDEdisc}).    At  the   same  time,   the  formal   continuum
Eq. (\ref{SPDE}) is  a useful device in order to  obtain a closed form
approximation of  the objects appearing in  (\ref{eq:SDEdisc}) such as
the mobility.

The microscopic view  does not give information about  the actual form
of  the diffusion  matrix and  free energy  functions and  we have  to
\textit{model} these  objects.  In the  present work, we  have modeled
the   bare   free  energy   functional   $\hat{\cal   F}[c]$  with   a
Ginzburg-Landau form.  We  have shown that the  continuum limit exists
in  1D for  a  sequence of  SDE of  the  form (\ref{eq:SDEdisc})  with
increasing number  of node points  by looking at the  structure factor
and  probability  distribution  of  particles in  a  region  of  fixed
extension.  

Although  simulation   results  have   been  presented  for   1D,  the
methodology  is  general  and  applicable,  in  principle,  to  higher
dimensions.   However, some  caution is  required in  $D>1$. From  the
results  on the  Gaussian model,  for which  the analytic  correlation
function  $\langle   \phi({\bf  r})\phi({\bf  r}')\rangle  $   can  be
explicitly  computed, we  know that  the point-wise  variance $\langle
\phi^2({\bf r})\rangle  $ diverges in  $D>1$. This quantity  gives the
average amplitude  of the field at  a given point.  This  implies that
the  fields $\phi({\bf  r})$  are extremely  irregular  and should  be
rather   understood  as   distributions   in  $D>1$   \cite{Ryser2012,
  HairerReview,   Hairer2012,  Hairer2014}.    However,  distributions
cannot  be  multiplied and  a  quartic  term  in  the free  energy  is
ill-defined.  This  means that the GL  model is ill defined  in $D>1$.
If one naively discretizes the corresponding ill-posed nonlinear SPDE,
pathological behavior  will be  observed in  $D>1$.  For  example, the
number of  particles in a  given \textit{finite} region of  space have
fluctuations that  depend on the  lattice spacing, which  is obviously
nonphysical.   There are  two  fundamentally  different approaches  to
address this problem.

On  one hand,  in  stochastic  field theories  as  those appearing  in
Quantum Field  Theory the governing  SPDE is postulated  from symmetry
arguments  without further  connection to  any other  more microscopic
description of the  system. In this case, the SPDE  is the fundamental
equation  and  one needs  to  modify  the  action (free  energy)  with
\textit{counterterms} that depend on the  lattice spacing in order for
the    final   theory    to    have   a    proper   continuum    limit
\cite{Benzi1989,Bettencourt1999,Gagne2000,Lythe2001,Cassol-Seewald2012}.

On the  other hand, when a  ``more fundamental'' theory exists,  as is
the  case in  colloidal suspension  where the  fundamental theory  are
Hamilton's equations, the  requirement of having a  continuum limit is
desirable   but  not   essential.    For   example,  the   microscopic
underpinning of the SPDE with a  free energy functional of the van der
Waals form shows that  such a model makes sense only  for a given cell
size (large  enough to  contain many particles,  small enough  for the
attractive  part  of  the  potential  to be  treated  in  mean  field)
\cite{Kampen1964}.   Indeed, if  the cell  size  was to  be taken  too
large, larger than droplet sizes, it would not be able to discriminate
between liquid and  vapor phases and the free energy  functional to be
used in  that case would  need to be  different from the  usual square
gradient van der Waals free energy functional.  It makes
no sense to  look at the mathematical continuum limit  of the discrete
SPDE,  while  the  discrete  SPDE   still  has  a  physically  sounded
foundation.

The finite element methodology presented may be extended to other SPDE
like  those  appearing  in the  Landau-Lifshitz  Navier-Stokes  (LLNS)
equations. For a  compressible theory, the free  energy function plays
essentially  the same  role as  in the  present theory.   In LLNS  one
usually  chooses a  free energy  which is  Gaussian. We  should expect
similar ultraviolet  catastrophic behavior  as in the  present simpler
non-linear diffusion.  However, the  Gaussian theory should still give
correct macroscopic observables like the amplitude of the fluctuations
of  the number  of particles  in  a \textit{finite}  region of  space.
While the  equilibrium properties  in the Gaussian  model do  not have
pathological  behavior,   the  convective  terms  in   the  equations,
involving   non-linear  terms,   require   a  careful   regularization
\cite{DiffusionJSTAT}.

\acknowledgements   

We  appreciate  useful  discussion  with Eric  Vanden-Eijnden,  Pedro
Tarazona, Mike  Cates, and  Marc Mel\'endez.   Discussion at  an early
stage of this work with Daniel Duque are greatly appreciated. A. Donev
was supported in part by the Office of Science of the U.S.  Department
of  Energy  through Early  Career  award  DE-SC0008271.  Support  from
MINECO  under grants  FIS2010-22047-C05-03 and { FIS2013-47350-C5-3-R} is acknowledged.

\appendix
\section{van der Waals and GL  free energy functionals}
\label{ap:free}
The van  der Waals  free energy  functional is  usually derived  for a
fluid  system  interacting with  a  pair-wise  potential that  can  be
separated into  a short  range repulsive  hard core  and a  long range
attractive                       part                      $\phi(r)<0$
\cite{Kampen1964,Reichl1980,Espanol2001}. This functional    was
proposed to describe the phase transition between liquid and vapor, a
transition  that  can  also   be  observed  in  colloidal  suspensions
\cite{Tata1992}. By  using a cell  method, with cells large  enough to
contain many  particles (i.e. size  of the  cell much larger  than the
molecular  volume  $b$ but  small  enough  to  treat the  long  ranged
attractive  interaction  in  mean   field),  van  Kampen  derived  the
following free energy functional, written by him in a continuum notation
 \begin{equation} 
{\cal F}[c]=\int d{\bf r} \left[f_0(c({\bf r}),T)
+\frac{1}{2}\int c({\bf r})c({\bf r'}) \phi({\bf r}-{\bf r}')d{\bf r}d{\bf r}'\right]  
\label{sl} 
\end{equation} 
where the attractive part is treated in mean field and the short range
part  of the  potential  produces the  local contribution  $f_0(c({\bf
  r}),T)$.  Under the  assumption  that the  density field  hardly
varies in the range of  the attractive potential, we may Taylor expand
$c({\bf  r}') =c({\bf r})+({\bf  r}'-{\bf r})\boldsymbol{\nabla}c({\bf
  r})+\cdots$ with the result
\begin{align}
\frac{1}{2}\int c({\bf r}) \phi({\bf r}-{\bf r}')c({\bf r'})d{\bf r}d{\bf r}'=& 
-a\int c({\bf r})^2d{\bf r}
\nonumber\\
&+\omega_2\int(\boldsymbol{\nabla} c({\bf r}))^2d{\bf r} 
\label{foupet2} 
\end{align} 
where we have defined 
\begin{eqnarray} 
a&\equiv&-\frac{1}{2}\int \phi(r)d{\bf r} 
\nonumber\\ 
\omega_2&\equiv&-\frac{1}{2}\int d{\bf r}{\bf r}^2\phi(r)
\label{omegas} 
\end{eqnarray} 
Note that $a>0 ,\omega_2>0$ for purely attractive potentials $\phi({\bf r})$. 
The free energy functional (\ref{sl}) becomes
 \begin{equation} 
{\cal F}^{\rm vdW}[c({\bf r})]=\int d{\bf r} 
\left[f(c({\bf r}),T)
+\omega_2(\boldsymbol{\nabla} c({\bf r}))^2\right]  
\label{gl1} 
\end{equation} 
where $f(c({\bf r}),T)=f_0(c({\bf r}),T)-ac({\bf r})^2$.
For a van der Waals gas, the free energy density is given by
\begin{equation}
f(c,T) = k_BTc\left[\ln\left(\frac{c\Lambda^D(T)}{1-cb}\right)-1\right]-ac^2
\label{fnt0}
\end{equation}
The  constants $a,b$ are  the attraction
parameter and the excluded  volume, respectively. The thermal wavelength is given by
$\Lambda(T) = h(2\pi m_0 k_BT)^{-1/2}$.

The Van der Waals gas is characterized by two critical parameters,
$T_c$ and $c_c$, obtained through the first and second derivatives
of the pressure, which leads to
\begin{align}
k_B T_c &= \frac{8a}{27 b} \nonumber \\
c_c &= \frac{1}{3b}
\end{align}
The  free  energy  functional  is  an example of  a  general class of free  energy
functionals  known  as  the Square  Gradient  Approximation.  In
general,  the coefficient  of the  square  gradient term  in the  free
energy depends also on the concentration \cite{Lutsko2011}.

The van der  Waals free energy can be approximated by neglecting
high order terms in an expansion around a constant concentration field,
leading to the  Ginzburg-Landau functional for the free energy.
If  we  put $c({\bf
  r})=c_0+\delta  c({\bf  r})$ we may expand to fourth order in $\delta
c({\bf r})$
\begin{align}
{\cal F}[c({\bf r})] =& \int d{\bf r} 
\left\{ a_0+b\delta c({\bf r}) \right. \nonumber \\
&+ \frac{1}{2}\left[\left.f_0''(c,T)\right|_{c_0} -2a\right]\delta c({\bf r})^2
 + \frac{1}{3!}\left.f_0'''(c,T)\right|_{c_0}\delta c({\bf r})^3 \nonumber \\
& + \frac{1}{4!}\left.f_0''''(c,T)\right|_{c_0}\delta c({\bf r})^4 + \left.\omega_2(\boldsymbol{\nabla} \delta c({\bf r}))^2\right\}
\label{gl2} 
\end{align}
Any constant term is irrelevant in the free energy and we can omit the
constant  term  $a_0$.   The   linear  term  in  $\delta  c({\bf  r})$
disappears  because of  the normalization  of the  density  field that
ensures $\int  \delta c({\bf r})d{\bf  r}=0$. For simplicity,  in this
paper we will restrict ourselves to be near the critical density, $c_0=c_c$.

The derivatives of $f_0$ are, at the critical density $c_c$,
\begin{align}
f_0^{'}    &= k_B T \left( \ln \left(\frac{b \Lambda^D(T)}{2}\right) + \frac{1}{2}\right)\nonumber\\
f_0^{''}   &= k_B T \frac{27}{4} b \nonumber \\
f_0^{'''}  &= 0 \nonumber \\
f_0^{''''} &= k_B T \frac{729}{8} b^3
\end{align}
so the free energy functional is obtained as
\begin{widetext}
\begin{align}
{\cal F}^{{\rm (GL)}}[ c({\bf r})] &= \frac{3}{8} \frac{k_B T}{b} \int
d{\bf r}\, \left\lbrace \frac{1}{8} \phi({\bf r})^4 + 
\left(1-\frac{k_BT_c}{k_BT} \right)\phi({\bf r})^2 + \sigma^2
\frac{k_B T_c}{k_B  T} \left(\nabla \phi({\bf r})\right)^2\right\rbrace
\label{eq:GLFEfunctionalb}
\end{align}
where we have defined $\phi({\bf r}) = (c({\bf r}) - c_0)/c_0$ and
$\sigma^2 = \omega_2 / a$.

\end{widetext}
It is obvious that the  Ginzburg-Landau free energy functional is only
a  good approximation  to the  van  der Waals  free energy  functional
around  the critical point  for which  the concentration  profiles are
close to  the homogeneous profile.   Nevertheless, the GL  free energy
already  captures the  essential of  a  phase transition  and we  will
restrict ourselves to this simpler model.  Note that the van der Waals
model does not  allow to have values of  the concentration larger than
$1/b$ (one molecule  per molecular volume) nor smaller  than zero.  On
the other hand, the GL model allows for unbounded values of $\phi({\bf
  r})$.  If the temperature is  much larger than the critical one, the
GL model reduces the Gaussian model.

\color{black}
\section{Other approximations for non-linear terms in the free energy}
\label{ap:approx-nonlinear}
While the  mass and stiffness  matrices introduced in  Eqs. (\ref{ML})
are  routinely computed  in finite  element algorithms,  the four-node
mass in Eq.  (\ref{M4}) may be  a rather cumbersome object to compute,
particularly in dimensions higher than  one.  For this reason, we will
also explore models  in which the non-quadratic  local contribution of
the  free  energy  is approximated  in  the  same  way  as we  did  in
Eq. (\ref{approx}). It is instructive first to look at the simpler
case of the GA model. In this model the static correlation is given by
\begin{align}
  C({\bf r},{\bf r}')&=\frac{c_0^2}{r_0} \delta({\bf r}-{\bf r}')
\label{crr0}\end{align}
{ This  result   can  be   obtained  from  the   limit  $k_0\to   0$  of
Eq.  (\ref{phirr'1D})}.   Note  that   this  is  also  the  equilibrium
correlation  function  of  the   concentration  field  of  independent
Brownian particles and  is, so to speak, a  ``physical'' result.  From
the  correlation of  the fields  (\ref{crr0}) we  can now  compute the
correlation matrix of the discrete variables
\begin{align}
  c_\mu&=\int d{\bf r}\delta_{\mu}({\bf r}) c({\bf r})
\label{CGc}
\end{align}
with the result
\begin{align}
  \langle \delta c_\mu \delta c_\nu\rangle^{\rm eq} &
  =\frac{c_0^2}{r_0} \int d{\bf r}\delta_\mu({\bf r})\delta_\nu({\bf r})
  =\frac{c_0^2}{r_0} M^\delta_{\mu\nu}
\label{correq}
\end{align}
Observe that the correlations of  the discrete concentration field are
due  to  the  overlapping  of the  weight  functions  $\delta_\mu({\bf
  r})$. Eq. (\ref{correq}) comes directly from the physics inherent to
(\ref{crr0})  and  the  definition  of the  coarse  grained  variables
(\ref{CGc}).  This is also the  result obtained for the Gaussian model
when  we   use  the  definition   of  the  free  energy   function  in
Eq. (\ref{Fdef}).  Therefore, (\ref{Fdef}) is a sensible definition.

The  Gaussian free  energy \textit{function}  can be  obtained exactly
from the free energy \textit{functional} (\ref{GA}) because it is just
a  quadratic functional  (and  similarly  for the  GL  leading to  the
explicit four-node mass).  For arbitrary functional forms of $f(c)$ we
cannot  proceed  by  computing   explicitly  the  space  integrals  in
(\ref{Fdef})  and we  need an  approximation scheme.   A naive  scheme
would  be to  approximate the  free  energy density  evaluated at  the
interpolated  field by  the  interpolated values  of  the free  energy
density at the nodes, this is
\begin{align}
  f({\bf c}\esc\boldsymbol{\psi}({\bf r}))\approx\sum_\mu \psi_\mu({\bf r})f(c_\mu)
\label{fap}\end{align}
Indeed, at the nodes ${\bf  r}={\bf r}_\mu$ this is an exact identity,
whereas at other points, the  function is approximated. If we
insert (\ref{fap}) into (\ref{Fdef}) we arrive at
\begin{align}
  F({\bf c}) &=\sum_\mu {\cal V}_\mu f(c_{\mu})
\label{naive}
\end{align}
which is certainly a natural and intuitive discretization of the space
integral.  Unfortunately, this  approximation  leads  to purely  local
correlations of the discrete variables, this is
\begin{align}
  \langle \delta c_\mu\delta c_\nu\rangle&\propto \delta_{\mu\nu}
\end{align}
which  does not conform  to the  physical result  (\ref{correq}).
The above discrepancy is expected to produce errors at length 
scales comparable to cell size, while for larger length scales the 
approximation (\ref{naive}) may be sufficient. 
It   may  be   convenient,  though,   to  improve   the  approximation
(\ref{naive}) in  order to not  miss even small scales  features, that
may  be important  in  hybrid methods  coupling  finite elements  with
particles.  In many cases of practical interest, the local part of the
free energy functional is of the form
\begin{align}
  {\cal F}[c]&=\int d{\bf r}c({\bf r})g(c({\bf r}))
\label{calFloc2}
\end{align}
where $g(c)$ is  a free energy per particle.  The discrete free energy
function  now becomes
\begin{align}
  F({\bf c}) &=\int d{\bf r}{\bf c}\esc\boldsymbol{\psi}({\bf r})
g({\bf c}\esc\boldsymbol{\psi}({\bf r}))
\end{align}
We may now approximate the free energy density according to 
\begin{align}
  g({\bf c}\esc\boldsymbol{\psi}({\bf r}))\approx\sum_\mu \psi_\mu({\bf r})g(c_\mu)
\end{align}
The result of this approximation is the explicit discrete free energy 
\begin{align}
  F({\bf c}) &={\bf c}^T{\bf M}^\psi{\bf g}({\bf c})
\label{Fdisfin}
\end{align}
where  ${\bf  g}({\bf  c})=(g(c_1),\cdots,g(c_M))$.   When  $g(c)$  is
linear in  the concentration, this approximation  recovers the correct
form  of the  correlations  of the  discrete  concentration field,  as
opposed  to the  approximation  (\ref{fap}).  In  the  future we  will
explore  (\ref{Fdisfin})  in  the  context  of  nonlinear  free-energy
functionals such as those appearing in  the ideal gas or van der Waals
models.

   The  methodology  presented  allows one  to  treat  also
non-local free  energy functionals. These functionals  usually involve
a  smoothed density  profile  that renders  the functional  non-local
\cite{Hansen1986}. For the sake of  the illustration, we consider here
the exact 1D free energy \textit{functional} for hard rods proposed by
Percus \cite{Percus1988} which has the form
\begin{align}
  {\cal F}[c]&=k_BT\int dz c(x)\ln (1-\sigma \overline{c}(x))
\end{align}
where  $\sigma$  is the  length  of  the  hard  rod and  the  smoothed
concentration profile is
\begin{align}
  \overline{c}(x)&=\frac{1}{\sigma}\int_{x-\sigma}^xdy c(y)
\end{align}
which is the space average of  the concentration field over the length
of the  rod.  The  recipe (\ref{Fdef})  now leads  to a  discrete free
energy \textit{function} of the form
\begin{align}
  F({\bf c})&=k_BT \sum_\mu c_\mu\int dz\psi_\mu(x)\ln(1-\sigma \overline{\overline{c}}(x))
\label{NonLocF}
\end{align}
where we have introduced
\begin{align}
\overline{\overline{c}}(x)\approx\frac{1}{\sigma}\int_{x-\sigma}^xdy \sum_\nu \psi_\nu(y)c_\nu
=\sum_\nu \overline{\psi}_\nu(x)c_\nu
\label{c1}
\end{align}
where the last identity defines the function
\begin{align}
 \overline{\psi}_\nu(x)\equiv  \frac{1}{\sigma}\int_{x-\sigma}^xdy \psi_\nu(y)
\end{align}

However, Eq. (\ref{NonLocF}) is still intractable because the integral
needs to be done numerically every  time we update ${\bf c}$. For this
reason, we propose to approximate the smoothed density field according to
\begin{align}
  \overline{c}(x)&=\sum_\mu\overline{c}_\mu \psi_\mu(x)
\label{c2}
\end{align}
which  is  just  a  linear interpolation  with  suitable  coefficients
$\overline{c}_\mu$  for the  discretization  of  the smoothed  density
field. We  evaluate the  coefficients $\overline{c}_\mu$  by requiring
compatibility beteween  (\ref{c1}) and  (\ref{c2}). This leads  to the
condition
\begin{align}
  \sum_\mu\psi_\mu(x)\overline{c}_\mu&=\sum_\mu\overline{\psi}_\mu(x){c}_\mu
\end{align}
By   multiplying   both   sides   with   $\delta_\mu(x)$   and   using
(\ref{projection0})    leads     to    the    explicit     form    for
$\overline{{c}}_\mu$ in  terms of the discrete  concentration $c_\mu$,
this is
\begin{align}
  \overline{{c}}_\mu&=\sum_\nu S_{\mu\nu}{c}_\nu
\end{align}
where the smoothing matrix $S_{\mu\nu}$ is given by the geometric object
\begin{align}
  S_{\mu\nu}&=\int dz\delta_\mu(x)\phi_\nu(x)
\end{align}
that may be computed  easily in an explicit way.
As a final step we take the approximation
\begin{align}
  \ln(1-\sigma \overline{c}(x))&=\sum_\mu\psi_\mu(x)\ln(1-\sigma\overline{c}_\mu)
\end{align}
which is  a natural way  of approximating  a function by  a piece-wise
linear expression. After inserting this result into (\ref{NonLocF}) we
obtain the  discrete free energy  function as an explicit  function of
the     concentration    field     and    the     geometric    objects
$M^\psi_{\mu\nu},S_{\mu\nu}$
\begin{align}
  F({\bf c})&=k_BT\sum_{\mu\nu} c_\mu M^\psi_{\mu\nu} \ln \left(1-\sigma \sum_{\nu'}S_{\nu\nu'}c_{\nu'}\right)
\end{align}
where $M^\psi_{\mu\nu}$ is defined in (\ref{Mdelta}).

It is important to note, however, that the above derivation has implicitly assumed that the hydrodynamic
cells are sufficently small to allow to treat the concentration as smooth over the cell length. In many
practical cases of interest, such as for example, layering near a wall, there will be spatial variability 
on length scales comparable to the length of a rod, and the above non-local recipe would require using cells that
contain fewer than a single particle; in this case it is not sensible to also include thermal fluctuations in
the description and only the equations for the average can be studied.

\color{black}

\section{1D discretization}
\label{App:1D}
The discrete  model (\ref{eq:SDEdisc}),  (\ref{FGLdis0}) is  valid for
any  space dimension.  In  this section  we present  the  model in  1D
explicitly.  The nodes are  at positions $x_\mu$, $\mu=1,\cdots,M$ and
the basis function $\psi_\mu({\bf r})$ is given by
\begin{align}
\psi_\mu(x) &= \theta(x-x_{\mu-1}) \theta(x_\mu - x) 
  \frac{x-x_{\mu-1}}{{\cal V}_{\mu}^l} \nonumber \\
            &\phantom{=}
             + \theta(x-x_{\mu}) \theta(x_{\mu+1} - x)
  \frac{x_{\mu+1}-x}{{\cal V}_\mu^r}
  \label{eq:DelaunayPsi}
\end{align}
where $\theta(x)$  is the Heaviside  step function, ${\cal  V}_\mu^r =
x_{\mu+1}-x_{\mu}$   and   ${\cal   V}_\mu^l   =   x_{\mu}-x_{\mu-1}$.
Figure~{\ref{fig:delaunay1D}}  shows  the function  $\psi_\mu(x)$  for
three  neighbor  cells.   The  resulting mass  ${\bf M}^\psi_{\mu\nu}$
and stiffness ${\bf L}^\psi_{\mu\nu}$  defined in Eqs. (\ref{ML}) take
the form
\begin{align}
{\bf M}^\psi_{\mu\nu}&=
\left\lbrace
\begin{array}{cl}
\frac{1}{6} {\cal V}_{\mu}^l & \mathrm{iff} \, \nu = \mu-1 \\
\frac{1}{3} ({\cal V}_\mu^l + {\cal V}_\mu^r) & \mathrm{iff} \, \nu = \mu \\
\frac{1}{6} {\cal V}_{\mu}^r & \mathrm{iff} \, \nu = \mu+1 \\
0 & \mathrm{otherwise}  \\
\end{array}
\right.
\nonumber\\
{\bf L}^\psi_{\mu\nu}&=
\left\lbrace
\begin{array}{cl}
-\frac{1}{{\cal V}_\mu^l} & \mathrm{iff} \, \nu = \mu-1 \\
\frac{1}{{\cal V}_{\mu}^l}
+
\frac{1}{{\cal V}_{\mu}^r}
& \mathrm{iff} \, \nu =
\mu \\
-\frac{1}{{\cal V}_\mu^r} & \mathrm{iff} \, \nu = \mu+1 \\
0 & \mathrm{otherwise} \\
\end{array}
\right.
\label{MLpsi}
\end{align}
whereas the  four-node mass $M_{\mu\nu\mu'\nu'}$ introduced
in (\ref{M4}) has the following elements.  For a given $\mu$, the only
non-zero elements are those in which  the other indices take the value
$\mu$, $\mu+1$ or $\mu-1$.  This gives the following non-zero elements
\begin{align}
M_{\mu\mu\mu\mu}^{\psi}             
   &= \frac{1}{5} ({\cal V}_\mu^r + {\cal V}_\mu^l) \nonumber \\
M_{\mu\mu\mu(\mu+1)}^{\psi}              
   &= \frac{1}{20} {\cal V}_\mu^r \nonumber \\
M_{\mu\mu\mu(\mu-1)}^{\psi}              
   &= \frac{1}{20} {\cal V}_\mu^l \nonumber \\ 
M_{\mu\mu(\mu+1)(\mu+1)}^{\psi}        
   &= \frac{1}{30} {\cal V}_\mu^r  \nonumber \\  
M_{\mu\mu(\mu-1)(\mu-1)}^{\psi}        
   &= \frac{1}{30} {\cal V}_\mu^l \nonumber \\
M_{\mu(\mu+1)(\mu+1)(\mu+1)}^{\psi}  
  &= \frac{1}{20} {\cal V}_\mu^r  \nonumber \\ 
M_{\mu(\mu-1)(\mu-1)(\mu-1)}^{\psi} 
  &= \frac{1}{20} {\cal V}_\mu^l 
\label{4nodemass}
\end{align}
The quartic contribution to the free energy function can be written as
  \begin{align}
    F^4_\mu(\phi) &=  \left[
    \phi_\mu^4 M_{\mu\mu\mu\mu}^{\psi} \right. 
    \nonumber \\ &\phantom{=}
    + 3 \phi_{\mu}^3 \phi_{\mu-1} M_{\mu\mu\mu(\mu-1)}
    \nonumber \\ &\phantom{=}
    + 3 \phi_{\mu}^2 \phi_{\mu-1}^2 M_{\mu\mu(\mu-1)(\mu-1)}
    \nonumber \\ &\phantom{=}
    +   \phi_{\mu} \phi_{\mu-1}^3 M_{\mu(\mu-1)(\mu-1)(\mu-1)}
    \nonumber \\ &\phantom{=}
    + 3 \phi_{\mu}^3 \phi_{\mu+1} M_{\mu\mu\mu(\mu+1)}
    \nonumber \\ &\phantom{=}
    + 3 \phi_{\mu}^2 \phi_{\mu+1}^2 M_{\mu\mu(\mu+1)(\mu+1)}
    \nonumber \\ &\phantom{=}\left.
    +   \phi_{\mu} \phi_{\mu+1}^3 M_{\mu(\mu+1)(\mu+1)(\mu+1)}
  \right]
  \end{align}
\color{black}
  By  using (\ref{MLpsi})  and (\ref{4nodemass}),  the GL  free energy
  (\ref{FGLdis0}) becomes in 1D the following function
\begin{widetext}
\begin{align}
 F^{(GL)}(\phi) &=
 k_B T \sum_\mu \left\lbrace
    \frac{u_0}{4} F^{(4)}_\mu
    + \frac{1}{6}\frac{r_0}{2} \left[
        \phi_\mu \phi_{\mu+1} {\cal V}_{\mu}^r 
        +
        2 \phi_\mu^2 ({\cal V}_{\mu}^r + {\cal V}_{\mu}^l)
        +
        \phi_\mu \phi_{\mu-1} {\cal V}_{\mu}^l 
    \right] \right.
\nonumber\\
&     + \left.
     \frac{K}{2} \left[
         -\phi_{\mu} \phi_{\mu+1} \frac{1}{{\cal V}_{\mu}^r}
         +\phi_{\mu}^2
         \frac{{\cal V}_\mu^r + {\cal V}_\mu^l}{{\cal V}_\mu^r {\cal
         V}_\mu^l}
         -\phi_{\mu} \phi_{\mu-1} \frac{1}{{\cal V}_\mu^l} 
     \right]
  \right\rbrace
\label{eq:GLFEdiscrete}
\end{align}
where
\begin{align}
 F^{(4)}_{\mu} &=
 \frac{1}{20} \left(
    4 \phi_{\mu}^4                  
  + 3 \phi_{\mu}^3 \phi_{\mu+1}   
  + 2 \phi_{\mu}^2 \phi_{\mu+1}^2 
  +   \phi_{\mu}   \phi_{\mu+1}^3 
 \right) {\cal V}_\mu^r 
 + 
 \frac{1}{20} \left(
    4 \phi_{\mu}^4                  
  + 3 \phi_{\mu}^3 \phi_{\mu-1}   
  + 2 \phi_{\mu}^2 \phi_{\mu-1}^2 
  +   \phi_{\mu}   \phi_{\mu-1}^3 
 \right){\cal V}_\mu^l 
\end{align}
\subsubsection{Time discretization}
We discuss now the temporal integrator for the stochastic diffusion equation  (\ref{eq:SDEdisc})
\begin{align}
    \frac{d {\bf c}}{dt}(t)
    &=
    - {\bf D} \frac{\partial F({\bf c}(t))}{\partial {\bf c}} 
    +  {\bf K} \boldsymbol{\mathcal{W}}(t)   
\end{align}
For constant mobility the diffusion matrix in  Eq. (\ref{eq:Dcte}) is constant and independent of ${\bf c}$
\begin{align}
  {\bf  D} &=  \frac{D c_0}{k_B  T} {\bf  M}^\delta {\bf  L}^\psi {\bf M}^\delta
\end{align}
The noise (\ref{eq:finalnoise}) has the explicit form in 1D
\begin{align}
  \frac{d\tilde{\bf c}}{dt} &= {\bf K}\boldsymbol{\mathcal{W}}(t)=
\sqrt{2 D c_0} {\bf M}^{\delta} {\bf N}^{\psi}\boldsymbol{\mathcal{W}}(t)
\end{align}
where 
\begin{align}
  {\bf N}^{\psi} &=
    \left(
    \begin{array}{rrrrrr}
    0 &  1/\sqrt{{\cal V}_1^r} &             0 &  0 \cdots
      & -1/\sqrt{{\cal V}_1^l} \\
        -1/\sqrt{{\cal V}_2^l}  &             0 &  
         1/\sqrt{{\cal V}_2^r}  &  0 \cdots & 0 \\
    0 & -1/\sqrt{{\cal V}_3^l} &             0 &
         1/\sqrt{{\cal V}_3^r}  \cdots & 0 \\
          &   &   &   \ddots &   \\
    \end{array}
    \right)
\end{align}
and the vector  $\boldsymbol{\mathcal{W}}$ is a collection of M independent white-noise processes. Note  that for periodic systems in 1D  the number  of elements (which  are segments
between the nodes) coincide with the number of nodes.

For the free energy function (\ref{eq:GLFEdiscrete}) the SDE becomes
\begin{align}
    \frac{d {\bf c}}{dt}
    &=
    - \frac{D}{c_0} 
  \left(
     r_0 {\bf M}^\delta {\bf L}^\psi 
    + K {\bf M}^\delta {\bf L}^\psi {\bf M}^\delta {\bf L}^\psi
  \right)
    {\bf c}
    - D u_0 {\bf M}^\delta {\bf L}^\psi {\bf M}^\delta
    {\bf c}{'}    
    + \sqrt{2 Dc_0} {\bf M}^{\delta} {\bf N}^{\psi}\boldsymbol{\mathcal{W}}(t)
\nonumber\\
&\equiv {\bf L} {\bf c} + {\bf g}({\bf c}) 
    +  {\bf K} \boldsymbol{\mathcal{W}}\end{align}
where we have introduced 
\begin{align}
  {\bf L} &=   - \frac{D}{c_0} 
  \left(
     r_0 {\bf M}^\delta {\bf L}^\psi 
    + K {\bf M}^\delta {\bf L}^\psi {\bf M}^\delta {\bf L}^\psi 
  \right) \nonumber
 \\
{\bf g}({\bf c}) &=
    - D u_0 {\bf M}^\delta {\bf L}^\psi {\bf M}^\delta
    {\bf c}{'}
\nonumber\\
  c_\mu' &=
 \frac{1}{20} \left(
    4 \phi_{\mu}^3                  
  + 3 \phi_{\mu}^2 \phi_{\mu+1}   
  + 2 \phi_{\mu}^1 \phi_{\mu+1}^2 
  +                \phi_{\mu+1}^3 
 \right) {\cal V}_\mu^r 
 + 
 \frac{1}{20} \left(
    4 \phi_{\mu}^3                  
  + 3 \phi_{\mu}^2 \phi_{\mu-1}   
  + 2 \phi_{\mu}^1 \phi_{\mu-1}^2 
  +                \phi_{\mu-1}^3 
 \right){\cal V}_\mu^l 
\end{align}
We  recognize  a  term  ${\bf  L}{\bf  c}$  which  is  linear  in  the
concentration and  a non-linear term  $g({\bf c})$ due to  the quartic
contribution  to  the  free  energy.    The  linear  term  is  just  a
discretization of a  diffusion term combining second  and fourth order
derivatives. In  order to avoid  instabilities and  to be able  to use
large time  steps, the  linear term is  treated implicitly,  while the
non-linear term is treated explicitly.  By following the semi-implicit
trapezoidal       predictor-corrector       scheme       of       Refs.
\cite{PhysRevE.87.033302,MultiscaleIntegrators}   we  may   write  the
following temporal integrator scheme
\begin{align}
    \left(
      {\bf M}^\psi + \frac{\Delta t}{2} \frac{D}{c_0} 
        \left(
              r_0 {\bf L}^\psi
            + K {\bf L}^\psi {\bf M}^\delta {\bf L}^\psi 
        \right)
    \right) \tilde{\bf c}^{n+1} &=
    \left(
      {\bf M}^\psi - \frac{\Delta t}{2} \frac{D}{c_0}
        \left(
              r_0 {\bf L}^\psi
            + K {\bf L}^\psi {\bf M}^\delta {\bf L}^\psi 
        \right)
    \right) {\bf c}^{n} \nonumber \\
    &\phantom{=}
    - \Delta t D u_0 {\bf L}^\psi {\bf M}^\delta 
    {\bf c}{'}^n
    + \sqrt{2Dc_0 \Delta t} {\bf N}^{\psi} {\bf W}^n \nonumber
    \\
    \left(
      {\bf M}^\psi + \frac{\Delta t}{2} \frac{D}{c_0}
        \left(
              r_0 {\bf L}^\psi
            + K {\bf L}^\psi {\bf M}^\delta {\bf L}^\psi 
        \right)
    \right) {\bf c}^{n+1} &=
    \left(
      {\bf M}^\psi - \frac{\Delta t}{2} \frac{D}{c_0}
        \left(
              r_0 {\bf L}^\psi
            + K {\bf L}^\psi {\bf M}^\delta {\bf L}^\psi 
        \right)
    \right) {\bf c}^{n} 
    \nonumber \\ &\phantom{=}
    - D u_0 {\bf L}^\psi {\bf M}^\delta 
    \left(\frac{{\bf c}{'}^n+ \tilde{\bf c}{'}^{n+1}}{2}\right)
    + \sqrt{2Dc_0 \Delta t} {\bf N}^{\psi} {\bf W}^n 
    \label{eq:NumIntegrator}
  \end{align}
  Here, ${\bf W}^n$  denotes a collection of  standard Gaussian random
  numbers  generated independently  at  each time  step.  For  regular
  lattices,  the set  of Eqs.~(\ref{eq:NumIntegrator})  can be  computed
  efficiently by using Fast Fourier  Transform (FFT) in any dimension.
  Indeed,   as   we    show   in   Appendix~\ref{app:structure}   (see
  Eq.~(\ref{eigenm}) obtained from Eq.~(\ref{eq:IntSCHEME})), we
  may diagonalize  the matrices in  Fourier space, obtaining a  set of
  uncoupled SODE.

  For non uniform meshes we cannot use the FFT procedure, but we still
  have  a  set of  tridiagonal  matrices  ${\bf M}^{\psi}$  and  ${\bf
    L}^{\psi}$  (see  Eq.~(\ref{MLpsi})).   The  right  hand  side  of
  (\ref{eq:NumIntegrator}) can be solved  using a specialized backward
  substitution, which is by far more efficient than operating with the
  dense  matrix  ${\bf  M}^{\delta}=[{\bf  M}^\psi]^{-1}$.  Since  the
  matrix on  the left hand  side in Eq. (\ref{eq:NumIntegrator})  is a
  constant Hermitian  positive-definite matrix,  it can  be decomposed
  with   a  Cholesky   factorization,   which  allows   us  to   solve
  Eq. (\ref{eq:NumIntegrator}) efficiently.

 \end{widetext}
\section{Structure factor for Gaussian models}
\label{app:structure}
In this appendix we present analytic results for the Gaussian model in
both continuum  and discrete  settings.  The main  result is  that the
numerical algorithm  closely matches not only  the infinite resolution
limit, but more importantly, it matches closely the predictions of the
continuum theory  for the  fluctuations in finite  resolution discrete
lattices.
\subsection{Static structure factor from the continuum}
The equilibrium  correlation of the fluctuations  of the concentration
is translationally invariant
\begin{align}
    \left\langle \delta c({\bf r},0) \delta c({\bf r}',0)\right\rangle = S({\bf r} - {\bf r}')
\label{covariance}
\end{align}
for some  function $S({\bf r})$.   Due to translation  invariance, the
Fourier transform $S({\bf k})$ of  the function $S({\bf r})$, known as
the static structure factor, is given by
\begin{align}
    S({\bf k}) &= 
    \left\langle \delta c({\bf k},0) \delta c(-{\bf k},0)\right\rangle 
\end{align}

Note that $S({\bf k}) =  c_0^2 S_{\phi}({\bf k})$, with $S_{\phi}({\bf
  k}) =  \left\langle \phi({\bf k},0)  \phi(-{\bf k},0)\right\rangle$,
and $\phi({\bf r})$ is the  relative fluctuations of the concentration
field.  The  static structure factor  is the Fourier transform  of the
second    moments   of    the    functional   probability    $P[c]\sim
\exp\{-\frac{1}{k_BT} {\cal  F}[c]\}$.  For a Gaussian  probability we
may  compute the  second  moments in  a  straightforward manner.   The
probability   functional    (\ref{Peq})   with   the    model   ${\cal
  F}^{GA+\sigma}[c]$ given  in (\ref{GA})  can be written  in operator
notation as
\begin{align}
  P^{\rm eq}[c]
&\propto\exp\left\{-\frac{1}{2}
\int d{\bf r}\int d{\bf r}'\, \phi({\bf r}){\cal L}({\bf r},{\bf r}')
\phi({\bf r}')
\right\}
\label{PeqGA}
\end{align}
where we have introduced the kernel
\begin{align}
  {\cal L}({\bf r},{\bf r}') &=r_0\delta({\bf r}-{\bf r}')-
K\nabla^2\delta({\bf r}-{\bf r}')
\end{align}
The covariance of the Gaussian probability functional is given by
\begin{align}
  \langle \phi({\bf r})\phi({\bf r'})\rangle &={\cal L}^{-1}({\bf r},{\bf r}')
\end{align}
where ${\cal L}^{-1}({\bf r},{\bf r}')$ is the inverse of the operator ${\cal L}({\bf r},{\bf r}')$, satisfying
\begin{align}
  \int d{\bf r}'{\cal L}({\bf r},{\bf r}'){\cal L}^{-1}({\bf r}',{\bf r}'')=\delta({\bf r}-{\bf r}'')
\end{align}
By inserting the form of the operator ${\cal L}({\bf r},{\bf r}')$ one
recognizes  that the  inverse operator  is just  the  Green's function
${\cal L}^{-1}({\bf r},{\bf r}')=S_\phi({\bf r}-{\bf r}')$, which satisfies
\begin{align}
  r_0S_\phi({\bf r}-{\bf r}')-K\nabla^2S_\phi({\bf r}-{\bf r}')=\delta({\bf r}-{\bf r}')
\label{Green}
\end{align}
The  solution  of  this  equation  is obtained  by  going  to  Fourier
space. We introduce the Fourier transform
\begin{align}
\hat{S}_\phi({\bf k})&=\int d^D{\bf r}e^{-i{\bf k}\esc{\bf r}}S_\phi({\bf r})
\nonumber\\
S_\phi({\bf r})&=\int \frac{d^D{\bf k}}{(2\pi)^D}e^{i{\bf k}\esc{\bf r}}\hat{S}_\phi({\bf k})
\end{align}
In Fourier space, (\ref{Green}) becomes
\begin{align}
  r_0\hat{S}_\phi({\bf k})+Kk^2\hat{S}_\phi({\bf k})=1
\label{GreenFourier}
\end{align}
which gives
\begin{align}
\hat{S}_\phi({\bf k})&=\frac{1}{r_0}\frac{1}{1+k^2/k_0^2}
\label{Gk}
\end{align}
where $k_0^2=r_0/K$. Therefore, the Green function is
\begin{align}
  S_\phi({\bf r})&=\int \frac{d^D{\bf k}}{(2\pi)^D}e^{i{\bf k}\esc{\bf r}}  \frac{1}{r_0}\frac{1}{1+k^2/k_0^2}
\end{align}
and the covariance, or correlation function, is given by
\begin{align}
\langle \phi({\bf r})\phi({\bf r'})\rangle &=
\int \frac{d^D{\bf k}}{(2\pi)^D}e^{i{\bf k}\esc({\bf r}-{\bf r}')}  \frac{1}{r_0}\frac{1}{1+k^2/k_0^2}
\label{phirr'}
\end{align}

For $D=1$  this correlation takes the form
\begin{align}
\langle \phi({\bf r})\phi({\bf r'})\rangle &=
\frac{k_0}{2r_0}e^{-k_0|r-r'|}
\label{phirr'1D}
\end{align}
For $D=2$ the result is
\begin{align}
\langle \phi({\bf r})\phi({\bf r'})\rangle 
&=
\frac{k^2_0}{4\pi r_0}K_0(k_0|{\bf r}-{\bf r}'|)
\label{phirr'2D}
\end{align}
where $K_0(x)$ is a Bessel function.
Finally, in $3D$ the result is 
\begin{align}
\langle \phi({\bf r})\phi({\bf r'})\rangle 
&=\frac{k_0^2}{4\pi r_0}\frac{e^{-k_0|{\bf r}-{\bf r}'|}}{|{\bf r}-{\bf r}'|}
\label{D15}
\end{align}
Note that  the quantity $\langle  \phi^2({\bf r})\rangle $  that gives
the  normalized fluctuations  of the  concentration field  at a  given
point of space does not diverge in $1D$ but it diverges  in
$2D$  and 3D, a  phenomenon  known  as the  ultraviolet
catastrophe.   This   means  that  the  point-wise   fluctuations  are
unbounded in  dimensions higher than one.   Any particular realization
of  the  field  will   be  extremely  rough.   

Nevertheless,  physical
observables like the  number of particles in a finite  region are well
behaved.   From a  physical  point  of view  this  quantity should  be
independent of  the resolution  used to  discretize the  problem.  The
number of particles in a region $V$ is given by
\begin{align}
  N_V&=\int_V d{\bf r} c({\bf r})
\end{align}
and the relative fluctuations are given by
\begin{align}
\phi_V&\equiv \frac{  N_V-Vc_0}{Vc_0}=\frac{1}{V}\int_V d{\bf r} \phi({\bf r})
\end{align}
The variance of this fluctuation is
\begin{align}
  \langle \phi^2_V\rangle &=
\frac{1}{V^2}\int_V d{\bf r} \int_V d{\bf r}' \langle \phi({\bf r})\phi({\bf r}')\rangle
\end{align}
This  quantity is  finite  for any  finite volume  but  as the  domain
shrinks to a point it diverges  in 2D logarithmically with the size of
the  domain, and  in 3D  inversely  with the  size of  the domain,  in
agreement with (\ref{D15}).

\color{black}

\subsection{Dynamic structure factor from the continuum}
In  this section  we  compute  the dynamic  structure  factor for  the
Gaussian  model.  Assume  a  constant  mobility $\Gamma=Dc_0/k_BT$  in
(\ref{SPDE})   with    the   model   ${\cal    F}^{GA+\sigma}[c]$   in
Eq. (\ref{GA}). The resulting SPDE is
\begin{align}
  \partial_t \delta c({\bf r},t) &=
D\frac{r_0}{c_0} \left( \nabla^2 \delta c({\bf r},t) -
  \frac{1}{k_0^2} \nabla^2 \nabla^2 \delta c \right)
\nonumber\\
&    + \sqrt{2Dc_0} \nabla \boldsymbol{\zeta}({\bf r},t)
\label{SPDEmucte}
\end{align}
where  we  have   introduced  $k_0^2=r_0/K$,  $\boldsymbol{\zeta}({\bf
  r},t)$ is  a white noise in  space and time, this  is, $\left\langle
  \boldsymbol{\zeta}({\bf r},t)\right\rangle  = 0$,  and $\left\langle
  \boldsymbol{\zeta}({\bf   r},t)    \boldsymbol{\zeta}({\bf   r}',t')
\right\rangle = \delta({\bf r} - {\bf r}')\delta(t-t')$.  Let us solve
the SPDE (\ref{SPDEmucte}) by Fourier transform
\begin{align}
  \partial_t \delta \hat{c} ({\bf k},t)
  &= - \frac{1}{\tau_k} \delta \hat{c} ({\bf k},t)
- i {\bf k} \sqrt{2 D c_0} \hat{\boldsymbol{\zeta}}({\bf k}, t) 
\label{SPDE_Fourier}
\end{align}
where we have introduced the relaxation time 
\begin{align}
  \tau_k &=  \left(  \frac{D}{c_0} r_0 \left(1 + \frac{k^2}{k_0^2}\right) k^2 \right)^{-1}
\end{align}
The Fourier transform of a white noise is also a white noise
which obeys the properties, $\left\langle \hat{\boldsymbol{\zeta}}({\bf
k},t)\right\rangle = 0$ and $\left\langle 
\hat{\boldsymbol{\zeta}}({\bf k},t)
\hat{\boldsymbol{\zeta}}({\bf k}',t')
\right\rangle = \delta({\bf k} + {\bf k}')\delta(t-t')$.

The linear equation (\ref{SPDE_Fourier}) has the explicit solution
\begin{align}
  \delta \hat{c}({\bf k},t) &= 
  \delta \hat{c} ({\bf k},0) \exp\left\lbrace 
  - \frac{t}{\tau_k} \right\rbrace
\nonumber\\
&  - i {\bf k} \sqrt{2 D c_0} 
  \int_0^t dt' \, e^{\frac{t-t'}{\tau_k}}\hat{\boldsymbol{\zeta}}(t')
\end{align}

By multiplying with  respect to the initial  condition $\delta \hat{c}
(-{\bf k},0)$ and  averaging with respect to  all possible equilibrium
realization of the initial condition we obtain
\begin{align}
\left\langle 
\delta \hat{c} ({\bf k}, t) \delta \hat{c} (-{\bf k},0)
\right\rangle
&=
S^c({\bf k}) \exp 
\left\lbrace
- \frac{t}{\tau_k}
\right\rbrace
\end{align}
where the  static structure factor is given in Eq. (\ref{Gk}).

\begin{widetext}
\section{The discrete static structure factor}

As a first step in order to obtain the discrete static structure
factor, we consider the integrator scheme (\ref{eq:NumIntegrator})
for the discrete density field ${\bf c}$ in the GA+$\sigma$ model,
given by
\begin{align}
    \left(
        {\bf M}^\psi + \frac{\Delta t}{2} \frac{D}{c_0} 
        \left(
              r_0 {\bf L}^\psi
            + K {\bf L}^\psi {\bf M}^\delta {\bf L}^\psi 
        \right)
    \right) {\bf c}^{n+1} &=
    \left(
        {\bf M}^\psi - \frac{\Delta t}{2} \frac{D}{c_0} 
        \left(
              r_0 {\bf L}^\psi
            + K {\bf L}^\psi {\bf M}^\delta {\bf L}^\psi 
        \right)
    \right) {\bf c}^{n} 
    + \sqrt{2 D c_0\Delta t}  {\bf N}^{\psi} \boldsymbol{\mathcal{W}}^n
     \label{eq:IntSCHEME}
\end{align}
which is a  matricial SODE. We seek for a  transformation that renders
the system  diagonal, leading to  a set  of uncoupled SODE  trivial to
solve  for each  $c_{\mu}$ value.  The $M$  vectors ${\bf  v}(m)=\{e^{i
  \frac{2  \pi}{L} m  x_\mu},\mu=0,\cdots M-1\}$  for $m=0,\cdots,M-1$
diagonalize simultaneously the three matrices involved in a 1D regular
lattice of spacing $a$, this is
\begin{align}
{\bf M}^{\psi} {\bf v}(m) 
  &= \frac{a}{3}\left[2+\cos\left(\frac{2 \pi m a}{L}\right)\right] {\bf v}(m) 
   = \widehat{m}(m) {\bf v}(m)\nonumber \\
{\bf L}^{\psi} {\bf v}(m)  
  &= \frac{2}{a}\left[1-\cos\left(\frac{2 \pi m a}{L}\right)\right] {\bf v}(m) 
   = \widehat{l}(m) {\bf v}(m) \nonumber \\
{\bf N}^{\psi} {\bf v}(m)  
  &= \frac{2}{\sqrt{a}}\sin\left( \frac{\pi m a}{L} \right) {\bf v}(m)
   = \sqrt{\widehat{l}}
   = \widehat{n}(m) {\bf v}(m)
\label{eigenm}
\end{align}
These equations define the eigenvalues of the problem that will be used below.

The (continuum) structure factor is defined as the Fourier transform of the static correlation function
\begin{align}
\hat{S}(k)&=\int_\infty^\infty dr \langle \delta c(0)\delta c(r)\rangle e^{-ikr}
\end{align}
Eq. (\ref{Gk}) shows that 
\begin{align}
\hat{S}(k)&= c_0^2 S_{\phi}(k) = \frac{c_0^2}{r_0}\frac{1}{1+k^2/k_0^2}
\label{Sk1}
\end{align}
In  order to  compare  numerical results  from  simulations with  this
theoretical results, it is necessary  to take into account the effects
of the discretization in the  continuum expression.  In this appendix,
we obtain from (\ref{Sk1}) the corresponding discrete structure factor
predicted  from  the continuum  theory   taking into  account  the
lattice spacing effects.  Explicit results are presented for a regular periodic
1D lattice.   

We introduce  the Fourier  series representation  of the
continuum concentration field
\begin{align}
 {c}(r,t) &= \sum_{k} \hat{c}(k,t) e^{i  k \cdot r} 
\end{align}
where the  sum is over  all those $k  = \frac{2 \pi  }{L}\kappa$, with
integer $\kappa\in \mathbb{Z}$.   The Fourier coefficients
are given by
\begin{align}
 \hat{c}(k,t) &= \frac{1}{L}\int_0^L c(r,t) e^{-i k r}dr
\label{eq:FFT}
\end{align}
Note   that   translation   invariance  $\langle   \delta   c(r)\delta
c(r')\rangle=S(r-r')$ implies that
\begin{align}
  \langle \delta c(k)\delta c({k'})\rangle &=\delta_{k,-k'}S(k)
\end{align}
where $S(k)$ are the Fourier coefficients of $S(r)$,
\begin{align}
  S(k)&\equiv\frac{1}{L}\hat{S}(k)
\end{align}
and we  are abusing  notation and  understand $\delta_{k,-k'}$  as the
Kroenecker   delta   $\delta_{\kappa,-\kappa'}$   for   the   integers
$\kappa,\kappa'$                    corresponding                   to
$k=\frac{2\pi}{L}\kappa,k'=\frac{2\pi}{L}\kappa'$.

\label{SktSk}.

We express the second  moments of  the probability $P^{\rm  eq}({\bf c})$
in terms of the continuum structure factor $S(k)$ obtained above
\begin{align}
    \left\langle \delta c_\mu \delta c_{\nu}\right\rangle &=
    \int dr \delta_{\mu}(r) 
    \int dr'\delta_{\nu}(r')  
    \left\langle \delta c(r,0) \delta c(r',0)\right\rangle
\label{cmucnu}
\end{align}
where we have defined the Fourier transform of the basis function
\begin{align}
     \hat{\delta}_{\nu}(k) &\equiv \int d{r}  \delta_{\nu}({r})    e^{-ikr}
\label{deltak}
\end{align}

From the linear relationship 
between         basis        functions         $\delta_\mu(r)=\sum_\nu
M^\delta_{\mu\nu}\psi_\nu(r)$ we obtain directly the explicit
functional form for the Fourier transform of the basis function
\begin{align}
   \hat{\delta}_{\mu}(k) &=     \frac{\hat{\psi}_\mu(k)}{\widehat{m}(k)}
=\frac{\hat{\psi}_0(k)}{\widehat{m}(k)}e^{-i k r_{\mu}}
 \quad\quad\quad\mbox{where}\quad\quad k=\frac{2\pi}{L}\kappa
\label{deltamuk}
\end{align}
with the corresponding Fourier  transform of  the basis function  $\psi_\nu(r)$ for  a 1D
regular grid of lattice spacing $a$ given by
\begin{align}
 \hat{\psi}_{\nu}(k) &=  \int dr\, \psi_{\nu}(r) e^{-i kr}
 =  a\mathrm{sinc}^2\left(\frac{ka}{2}\right)e^{-i k r_{\nu}}
=\hat{\psi}_0(k)e^{-i kr_\nu}
\label{psifou}
\end{align}

Note  that  for  $k\to  0$, we  have  $\hat{\delta}_\mu(0)=1$,  which,  from
(\ref{deltamuk}) and (\ref{lc}) is what it should be.

Eq. (\ref{cmucnu}) gives the covariance  of the discrete field in real
space, in terms of the structure  factor, but we are interested in the
covariances  of  the  discrete   Fourier  transform  of  the  discrete
field.  To this  end, we  introduce the  discrete Fourier  transform $
\hat{c}_m$ with $m=0,M-1$ of  the discrete concentration field $c_\mu$
according to
\begin{align}
  \hat{c}_m&=\frac{1}{M}\sum_{\mu=0}^{M-1} e^{-i\frac{2\pi}{L} mr_\mu}c_\mu
&&  c_\mu=\sum_{m=0}^{M-1}e^{i\frac{2\pi}{L} mr_\mu}\hat{c}_m
\label{DFFT}\end{align}
We define the  discrete static  structure factor $\hat{S}^c(k_m)$ as the
covariance of the discrete Fourier components $\hat{c}_m$
\begin{align}
\hat{S}^c(k_m)&\equiv L\left\langle \delta \hat{c}_m \delta \hat{c}^*_{m} \right\rangle
\nonumber\\
&    =\frac{L}{M^2}
    \sum_{\mu,\nu} e^{-i \frac{2 \pi}{L} m r_\mu } 
     e^{i \frac{2 \pi}{L} m r_\nu} 
    \left\langle {\delta c}_\mu {\delta c}_{\nu} \right\rangle
=\frac{L}{M^2} \sum_{\mu,\nu} e^{-i \frac{2 \pi}{L} m r_\mu } 
     e^{i\frac{2 \pi}{L} m r_\nu } 
\sum_{k} S(k)  \hat{\delta}_{\mu}(k)  \hat{\delta}_{\nu}(-k)
\nonumber\\
&=\sum_{ k} \hat{S}(k) 
    \overline{\delta}_{m }(k)
    \overline{\delta}_{-m}(-k)
\label{k1}
\end{align}
where $k_m=\frac{2\pi}{L}m$ and we have introduced the doubly Fourier transformed basis function
\begin{align}
  \overline{\delta}_{m}(k) &\equiv  
\frac{1}{M} \sum_{\mu} e^{-i \frac{2 \pi}{L} m r_\mu } 
  \hat{\delta}_{\mu}(k)=\frac{\hat{\psi}_0(k)}{\widehat{m}(k)}
\frac{1}{M}\sum_{\mu} e^{-i \frac{2 \pi}{L} m r_\mu }  e^{i k r_\mu} 
=\frac{\hat{\psi}_0(k)}{\widehat{m}(k)}
\sum_{\alpha\in\mathbb{Z}}\delta_{m,\kappa+\alpha M}
\label{doubly}
\end{align}
where $k=\frac{2\pi}{L}\kappa$ and  we have used the mathematical identity
\begin{align}
  \frac{1}{M} \sum_{\mu=0}^{M-1} e^{i \frac{2\pi}{L}mr_{\mu}}=\sum_{\alpha\in\mathbb{Z}}
\delta_{m,\alpha M}
\label{tricky}
\end{align}
In this way, we have
\begin{align}
\hat{S}^c(k_m)
&= \sum_k\sum_{\alpha\in\mathbb{Z}}\delta_{m,\kappa+\alpha M}
\sum_{\alpha'\in\mathbb{Z}}\delta_{m,\kappa+\alpha' M}\hat{S}(k)
\left[\frac{ \hat{\psi}_0(k)}{\widehat{m}(k)}\right]^2
\quad\quad\quad\quad \mbox{where} \quad k=\frac{2\pi}{L}m
\label{sc10}
\end{align}
Note that we have 
\begin{align}
  \sum_{\alpha'\in\mathbb{Z}}\delta_{m,\kappa+\alpha M}\delta_{m,\kappa+\alpha'M}=
  \sum_{\alpha'\in\mathbb{Z}}\delta_{m,\kappa+\alpha M}\underbrace{\delta_{\kappa+\alpha M,\kappa+\alpha'M}}_{\delta_{\alpha\alpha'}}=\delta_{m,\kappa+\alpha M}
\end{align}
and thus
\begin{align}
\hat{S}^c(k_m)
&=\sum_{\alpha\in\mathbb{Z}}\hat{S}\left(\frac{2\pi (m-\alpha M)}{L}\right)
\left[\frac{\hat{\psi}_0\left(\frac{2\pi (m-\alpha M)}{L}\right)}
{\widehat{m}\left(\frac{2\pi (m-\alpha M)}{L}\right)}\right]^2
\label{sc1}
\end{align}
After inserting  (\ref{Sk1}) into  (\ref{sc1}) we obtain  the discrete
structure factor for the GA+$\sigma$ model,
\begin{align}
\hat{S}^c(k)&=\frac{c_0^2}{r_0} \frac{9}{\left[2+\cos\left(ka\right)\right]^2}
\sum_{\alpha\in\mathbb{Z}}
\frac{\mathrm{sinc}^4\left(\frac{ka}{2}-\pi\alpha\right)}
{1+\left(\frac{k}{k_0}-\frac{2\pi \alpha }{k_0a}\right)^2}
\label{ScFIN-app}
\end{align}
where $k=\frac{2\pi}{L}m$.  Note that in  the limit of high resolution
$a=L/M\to 0$, the only term that  contributes in the sum over $\alpha$
is  $\alpha=0$.    In  this  limit,  the   discrete  structure  factor
(\ref{ScFIN-app}) converges  towards the continuum  limit (\ref{Sk1}).
Eq.   (\ref{ScFIN-app}) gives  the  the fluctuations  of the  discrete
concentration variables as obtained from  the continuum theory and our
definition of the coarse-grained variables.

In  the  limit   $k_0\to  \infty$  corresponding  to   the  GA  model,
(\ref{ScFIN-app}) becomes
\begin{align}
\hat{S}^c(k)
&=\frac{c_0^2}{r_0} \frac{3}{\left[2+\cos\left(ka\right)\right]}
\label{SckGA}
\end{align}
where $k=\frac{2\pi m }{L}$.
This is indeed the  correct result of the GA model as  it can be shown
by a  more direct  route.  In the  GA model, we  know that  the second
moments of the probability functional are given by
\begin{align}
  \langle \delta c(r)\delta c(r')\rangle&=\frac{c_0^2}{r_0}\delta(r-r') &&\to&&  \langle \delta c_\mu\delta c_\nu\rangle =\frac{c_0^2}{r_0}M^\delta_{\mu\nu}
\end{align}
where (\ref{natural}) has been used.
Next, by using Eq. (\ref{eigenm}) we obtain
\begin{align}
S^c(k)
    &
=\frac{c_0^2}{r_0}\frac{1}{M}
\frac{1}{\widehat{m}(k_n)} 
=\frac{c_0^2}{r_0}\frac{3}{\left[2+\cos\left(k a\right)\right] }
\label{D50}
\end{align}
which coincides with (\ref{SckGA}).

\section{{The discrete structure factor of the
numerical scheme}\label{app:DiscreteSKNum}}
The static structure  function (\ref{Sk1}) has been  computed from the
second moments of the probability  functional and can also be obtained
from  the  following  argument  that involves  the  continuum  dynamic
equation (\ref{SPDE_Fourier}).  In the limit $\Delta t \rightarrow 0$,
a simple  Euler integrator scheme  for
Eq.~(\ref{SPDE_Fourier}) gives
\begin{align}
  \delta \hat{c}^{n+1} &= 
  \delta \hat{c}^{n}
  - \frac{\Delta t}{\tau_k} \delta \hat{c}^n
  - i {\bf k} \sqrt{2Dc_0 \Delta t} \hat{\boldsymbol{\zeta}}^n
  \label{eq:EulerIntegrator}
\end{align}
If we multiply this equation by itself and average we obtain
\begin{align}
  \left\langle
  \delta \hat{c}^{n+1} \delta \hat{c}^{n+1} 
  \right\rangle
  &=
  \left( 1 - \frac{\Delta t}{\tau_k}\right)^2 
\left\langle
\delta \hat{c}^{n} \delta \hat{c}^{n}
\right\rangle
  + 2k^2 D c_0 \Delta t \nonumber \\
  &\simeq
  \left( 1 - 2 \frac{\Delta t}{\tau_k}\right) 
\left\langle
\delta \hat{c}^{n} \delta \hat{c}^{n}
\right\rangle
  + 2k^2   D c_0 \Delta t
\end{align}
where we have neglected terms of order $(\Delta t)^2$. At equilibrium  
$\left\langle
\delta \hat{c}^{n+1} \delta \hat{c}^{n+1}
\right\rangle = 
\left\langle
\delta \hat{c}^{n} \delta \hat{c}^{n}
\right\rangle = S^c({\bf k}) 
$, so that
\begin{align}
\hat{S}^c(k) &= k^2 D c_0 \tau_k =
  \frac{c_0^2}{r_0}
  \frac{1}{1 + k^2/k_0^2} 
  \label{eq:acf_k}
\end{align}
which coincides with (\ref{Sk1}).

The  same strategy  may  be  used to  compute  the discrete  structure
factor, by using the discrete time stepping scheme, and thus including
effects due to the finite time step \cite{Donev2010}.  In this way, one
may  obtain an  exact  prediction for  the  discrete structure  factor
$S^d(k)$ that is produced by the numerical code.  If the code is meant
to reproduce the  structure factor predicted by  the continuum theory,
we  should   have  $S^d(k)\approx   S^c(k)$  for   sufficiently  small
times.  The  only  difference  from   the  procedure  used  to  derive
Eq.~(\ref{eq:acf_k})  is  that  both  $k^2$ and  $\tau_k$  are  to  be
replaced by their corresponding discrete counterparts.

Namely, $\tau_k$ can simply be read from the fact that in the discrete
setting the integrator  scheme is given, in the  GA$+\sigma$ model, by
Eq.~(\ref{eq:NumIntegrator}) with no explicit  part. In Fourier space,
this equation  should give  us Eq.~(\ref{eq:EulerIntegrator}).   If we
equal      (\ref{eq:EulerIntegrator})     with      the     equivalent
({\ref{eq:NumIntegrator}}) in Fourier space we obtain
\begin{align}
  \tau_k &\mapsto \frac{c_0}{D r_0} \frac{\widehat{m}}{\widehat{l} +
\frac{1}{k_0^2} \frac{\widehat{l} \phantom{}^2}{\widehat{m}}} +
\mathcal{O}(\Delta t)
\end{align}
In the  same way,  the discrete  $k^2$ term can  be obtained  from the
covariance     of    the     noise     term     that    appears     in
(\ref{eq:EulerIntegrator}), which should coincide  with the noise term
in  (\ref{eq:NumIntegrator}),   giving  as   a  result   $k^2  \mapsto
\frac{\widehat{l}}{\widehat{m}\phantom{}^2}$, where  we have neglected
terms of order $\mathcal{O}(\Delta t)$.

In this way, the equivalent of Eq.~(\ref{eq:acf_k}) for the discrete
structure factor will be
\begin{align}
  \hat{S}^d(k) &= k^2 D c_0 \tau_k = \frac{c_0^2}{r_0}
  \frac{1}{\widehat{m}} \frac{1}{\widehat{l} +
  \frac{1}{k_0^2}\frac{\widehat{l}\phantom{}^2}{\widehat{m}}}
 =\frac{c_0^2 }{r_0}\frac{3}{\left[2+\cos\left(ka\right)\right]}\frac{1}{1+\frac{k^2}{k_0^2}\left(\frac{3 \mathrm{sinc}^2(ka/2)}{(2+\cos ka)}\right)}
\end{align}

\end{widetext}

\color{black}
\bibliographystyle{unsrt}

\begin{thebibliography}{10}

\bibitem{Kawasaki1994}
Kyozi Kawasaki.
\newblock {Stochastic model of slow dynamics in supercooled liquids and dense
  colloidal suspensions}.
\newblock {\em Physica A}, 208(1):35--64, July 1994.

\bibitem{Marconi1999}
UMB Marconi and P~Tarazona.
\newblock {Dynamic density functional theory of fluids}.
\newblock {\em J. Phys. Condens. Matter}, 12:A413, 2000.

\bibitem{Archer2004}
A~J Archer and R~Evans.
\newblock {Dynamical density functional theory and its application to spinodal
  decomposition.}
\newblock {\em J. Chem. Phys.}, 121(9):4246, September 2004.

\bibitem{Leonard2013}
T~Leonard, B~Lander, U~Seifert, and T~Speck.
\newblock {Stochastic thermodynamics of fluctuating density fields:
  Non-equilibrium free energy differences under coarse-graining.}
\newblock {\em J. Chem. Phys.}, 139(20):204109, November 2013.

\bibitem{Landau1959}
L.~D. Landau and E.~M. Lifshitz.
\newblock {\em Fluid Mechanics (First Edition)}.
\newblock Pergamon Press, 1959.

\bibitem{Archer2004a}
Andrew~J Archer and Markus Rauscher.
\newblock {Dynamical density functional theory for interacting Brownian
  particles: stochastic or deterministic?}
\newblock {\em J. Phys. A. Math. Gen.}, 37(40):9325--9333, October 2004.

\bibitem{Bray1994}
A.J. Bray.
\newblock {Theory of phase-ordering kinetics}.
\newblock {\em Adv. Phys.}, 43(3):357--459, June 1994.

\bibitem{Hohenberg1977}
PC~Hohenberg and BI~Halperin.
\newblock {Theory of dynamic critical phenomena}.
\newblock {\em Rev. Mod. Phys.}, 49(3):435, 1977.

\bibitem{Ryser2012}
Marc~D. Ryser, Nilima Nigam, and Paul~F. Tupper.
\newblock {On the well-posedness of the stochastic Allen–Cahn equation in two
  dimensions}.
\newblock {\em J. Comput. Phys.}, 231(6):2537--2550, March 2012.

\bibitem{Hairer2012}
Martin Hairer, Marc~Daniel Ryser, and Hendrik Weber.
\newblock {Triviality of the 2D stochastic Allen-Cahn equation}.
\newblock {\em Electron. J. Probab.}, 17:1--14, May 2012.

\bibitem{Benzi1989}
Roberto Benzi, Giovanni Jona-Lasinio, and Alfonso Sutera.
\newblock {Stochastically perturbed Landau-Ginzburg equations}.
\newblock {\em J. Stat. Phys.}, 55(3-4):505--522, May 1989.

\bibitem{Bettencourt1999}
Lu\'{\i}s Bettencourt, Salman Habib, and Grant Lythe.
\newblock {Controlling one-dimensional Langevin dynamics on the lattice}.
\newblock {\em Phys. Rev. D}, 60(10):105039, October 1999.

\bibitem{Gagne2000}
Cj~Gagne and M~Gleiser.
\newblock {Lattice-independent approach to thermal phase mixing}.
\newblock {\em Phys. Rev. E,}, 61(4 Pt A):3483--9, April 2000.

\bibitem{Lythe2001}
Grant Lythe and Salman Habib.
\newblock {Stochastic PDEs: convergence to the continuum?}
\newblock {\em Comput. Phys. Commun.}, 142(1-3):29--35, December 2001.

\bibitem{Cassol-Seewald2012}
N.~C. Cassol-Seewald, R.~L.~S. Farias, G.~Krein, and R.~S. {Marques De
  Carvalho}.
\newblock {Noise and Ultraviolet Divergences in Simulations of
  Ginzburg–Landau–Langevin Type of Equations}.
\newblock {\em Int. J. Mod. Phys. C}, 23(08):1240016, August 2012.

\bibitem{Hairer2014}
M~Hairer.
\newblock {A theory of regularity structures}.
\newblock {\em Invent. Math.}, -(October 2013):1--236, 2014.

\bibitem{HairerReview}
Martin Hairer.
\newblock {Introduction to Regularity Structures}.
\newblock {\em arXiv:1401.3014 [math.AP]}, 1(1):1, 2014.

\bibitem{DiffusionJSTAT}
Aleksandar Donev, Thomas~G Fai, and Eric Vanden-Eijnden.
\newblock {A reversible mesoscopic model of diffusion in liquids: from giant
  fluctuations to Fick’s law}.
\newblock {\em J. Stat. Mech. Theory Exp.}, 2014(4):P04004, April 2014.

\bibitem{Grabert1982}
H.~Grabert.
\newblock {\em Projection Operator Techniques in Nonequilibrium Statistical
  Mechanics}.
\newblock Springer Verlag, Berlin, 1982.

\bibitem{Saarloos1982}
W~Van Saarloos, D~Bedeaux, and P~Mazur.
\newblock {Non-linear hydrodynamic fluctuations around equilibrium}.
\newblock {\em Physica A}, pages 147--170, 1982.

\bibitem{Espanol1998a}
P.~Espa\~{n}ol.
\newblock {Stochastic differential equations for non-linear hydrodynamics}.
\newblock {\em Phys. A Stat. Mech. its Appl.}, 248(1):77--96, 1998.

\bibitem{Zubarev1983}
DN~Zubarev and VG~Morozov.
\newblock {Statistical mechanics of nonlinear hydrodynamic fluctuations}.
\newblock {\em Physica A}, pages 411--467, 1983.

\bibitem{Espanol2009}
P.~Espa\~{n}ol and I.~Z\'{u}\~{n}iga.
\newblock {On the definition of discrete hydrodynamic variables.}
\newblock {\em J. Chem. Phys.}, 131(16):164106, October 2009.

\bibitem{Espanol2009c}
P.~Espa\~{n}ol, J.G. Anero, and I.~Z\'{u}\~{n}iga.
\newblock {Microscopic derivation of discrete hydrodynamics.}
\newblock {\em J. Chem. Phys.}, 131(24):244117, December 2009.

\bibitem{DelaTorre2011}
J. A. de la Torre  and P.~Espa\~{n}ol.
\newblock {Coarse-graining Brownian motion: From particles to a discrete
diffusion equation}
\newblock {\em J. Chem. Phys.}, 135:114103, 2011. 

\bibitem{Garcia-Ojalvo1999}
J.~Garc\'{\i}a-Ojalvo and J.M. Sancho.
\newblock {\em Noise in Spatially Extended Systems}.
\newblock Springer Verlag, Berlin, 1999.

\bibitem{Donev2010}
A.~Donev Bell, E.~Vanden-Eijnden, A.~L. Garcia, and J.~B.
\newblock {On the Accuracy of Explicit Finite-Volume Schemes for Fluctuating
  Hydrodynamics}.
\newblock {\em Commun. Appl. Math. Comput. Sci.}, 5(2):149, 2010.

\bibitem{Walsh2005}
J.~B. Walsh.
\newblock {Finite Element Methods for Parabolic Stochastic PDE's}.
\newblock {\em Potential Anal.}, 23(1):1--43, August 2005.

\bibitem{Yan2005}
Yubin Yan.
\newblock {Galerkin Finite Element Methods for Stochastic Parabolic Partial
  Differential Equations}.
\newblock {\em SIAM J. Numer. Anal.}, 43(4):1363--1384, January 2005.

\bibitem{Krein2009}
G.~Krein, J.M. Machado, and A.O. Pereira.
\newblock {Dynamics of the deconfinement transition of quarks and gluons in a
  finite volume}.
\newblock {\em Comput. Phys. Commun.}, 180(4):564--573, April 2009.

\bibitem{Plunkett}
P~A~T Plunkett, J~O~N Hu, Chris Siefert, and Paul~J Atzberger.
\newblock {Spatially adaptive stochastic methods for fluid-structure
  interactions subject to thermal fluctuations in domains with complex
  geometries}.
\newblock {\em preprint}, pages 1--27, 2014.

\bibitem{Finlayson1972}
B.C Finlayson.
\newblock {\em The method of weighted residuals and variational principles,
  with application in fluid mechanics, heat and mass transfer, Volume 87}.
\newblock Academic Press, Inc, 1972.

\bibitem{Espanol2009f}
Pep Espa\~{n}ol, Jes\'{u}s~G Anero, and Ignacio Z\'{u}\~{n}iga.
\newblock {Microscopic derivation of discrete hydrodynamics.}
\newblock {\em J. Chem. Phys.}, 131(24):244117, December 2009.

\bibitem{Hansen1986}
J.-P. Hansen and I.R. McDonald.
\newblock {\em Theory of Simple Liquids}.
\newblock Academic Press, London, 1986.

\bibitem{Anero2013}
Jes\'{u}s~G Anero, Pep Espa\~{n}ol, and Pedro Tarazona.
\newblock {Functional thermo-dynamics: a generalization of dynamic density
  functional theory to non-isothermal situations.}
\newblock {\em J. Chem. Phys.}, 139(3):034106, July 2013.

\bibitem{Touchette2009}
Hugo Touchette.
\newblock {The large deviation approach to statistical mechanics}.
\newblock {\em Phys. Rep.}, 478(1-3):1--69, July 2009.

\bibitem{Donev2014}
Aleksandar Donev and Eric Vanden-Eijnden.
\newblock {Dynamic density functional theory with hydrodynamic interactions and
  fluctuations.}
\newblock {\em J. Chem. Phys.}, 140(23):234115, June 2014.

\bibitem{Troster2007a}
A.~Tr\"{o}ster.
\newblock {Coarse grained free energies with gradient corrections from Monte
  Carlo simulations in Fourier space}.
\newblock {\em Phys. Rev. B}, 76(1):1--4, July 2007.

\bibitem{Troster2007}
A~Tr\"{o}ster and C.~Dellago.
\newblock {Coarse Graining the $\phi$ 4 Model: Landau-Ginzburg Potentials from
  Computer Simulations}.
\newblock {\em Ferroelectrics}, 354(1):225--237, August 2007.

\bibitem{Kampen1964}
NG~Van Kampen.
\newblock {Condensation of a classical gas with long-range attraction}.
\newblock {\em Phys. Rev.}, A135:362, 1964.

\bibitem{Reichl1980}
L.~E. Reichl.
\newblock {\em A modern course in statistical physics}.
\newblock Univ. of Texas Press, Austin, 1980.

\bibitem{Espanol2001}
P.~Espa\~{n}ol.
\newblock {Thermohydrodynamics for a van der Waals fluid}.
\newblock {\em J. Chem. Phys.}, 115(12):5392, 2001.

\bibitem{Tata1992}
BVR Tata, M~Rajalakshmi, and AK~Arora.
\newblock {Vapor-liquid condensation in charged colloidal suspensions}.
\newblock {\em Phys. Rev. Lett.}, 69(26):3778--3782, 1992.

\bibitem{PhysRevE.87.033302}
Steven Delong, Boyce~E. Griffith, Eric Vanden-Eijnden, and Aleksandar Donev.
\newblock Temporal integrators for fluctuating hydrodynamics.
\newblock {\em Phys. Rev. E}, 87:033302, Mar 2013.

\bibitem{MultiscaleIntegrators}
S.~Delong, Y.~Sun, , B.~E. Griffith, E.~Vanden-Eijnden, and A.~Donev.
\newblock {Multiscale Temporal Integrators for Fluctuating Hydrodynamics}.
 
\newblock Phys. Rev. E 90:063312 (2014).
\color{black}
\bibitem{Lutsko2011}
James~F Lutsko.
\newblock {Density functional theory of inhomogeneous liquids. IV.
  Squared-gradient approximation and classical nucleation theory.}
\newblock {\em J. Chem. Phys.}, 134(16):164501, April 2011.

 
\bibitem{Percus1988}
JK~Percus.
\newblock {Free energy models for nonuniform classical fluids}.
\newblock {\em J. Stat. Phys.}, 52:1157--1178, 1988.

\end{thebibliography}

\end{document}